\documentclass{article}

\usepackage[colorlinks,linkcolor=blue,citecolor=blue,urlcolor=blue]{hyperref}
\usepackage{chemarrow}
\usepackage{color}
\usepackage{graphicx}
\definecolor{grey}{rgb}{0.9,0.9,0.9}
\usepackage[a4paper]{geometry}

\title{\bf Physical approaches to the dynamics of genetic circuits: A tutorial}

\author{{\bf Jordi Garcia-Ojalvo}\\
{\em Departament de F\'isica i Enginyeria Nuclear}\\
{\em Universitat
Polit\`ecnica de Catalunya}\\
{\em Rambla de Sant Nebridi s/n, 08222 Terrassa, Spain}\\
{\em Email:} \href{mailto:jordi.g.ojalvo@upc.edu}{jordi.g.ojalvo@upc.edu}}
\date{May 22nd, 2011}

\begin{document}

\maketitle

\begin{abstract}
Cellular behavior is governed by gene regulatory processes that are intrinsically 
dynamic and nonlinear, and are subject to non-negligible amounts of random fluctuations.
Such conditions are ubiquitous in physical systems, where they have been studied for decades using
the tools of statistical and nonlinear physics.
The goal of this review is to show how approaches traditionally used in physics can help in
reaching a systems-level understanding of living cells.
To that end, we present an overview
of the dynamical phenomena exhibited by genetic circuits and their functional significance.
We also describe the theoretical and experimental approaches that are being used
to unravel the relationship
between circuit structure and function in dynamical cellular processes under the influence
of noise, both at the single-cell level
and in cellular populations, where intercellular coupling plays an important role.\\

\noindent
To be published in {\em Contemporary Physics}.
\end{abstract}

\section{Introduction}

One of the main questions to be answered in the quest towards understanding life is how structure relates with function in living systems, in particular in cells. This question can be asked at many levels, from the microscopic scale of single proteins to the macroscopic level of complete organisms. A substantial amount of evidence has recently pointed to the relevance of a ``mesoscopic" description, at the level of \emph{networks} of interacting genes and proteins that coordinately govern most cellular processes \cite{Dixon:2009p463}.
This picture has relegated the notion of ``one gene, one function" (and frequently ``one disease") that guided much of genetics and molecular biology in most of the 20th century. Within that framework, countless experimental evidence
was gathered that revealed the identity of genes and proteins involved in diverse cellular functions.
Using that invaluable information, it is now time to
rephrase the question posed above in terms of finding the relationship between cellular function and the architecture of the underlying genetic networks. Two main factors make the solution of this problem difficult.

First, the number of proteins and genes involved in these gene regulatory networks is usually very large. This, together with the frequently nonlinear character of their interactions, generates a highly complex behavior, riddled with multiple coexisting phenomena and impossible to understand from the sum of the effects of the individual network elements. The situation is further complicated by the ubiquitous existence of heterogeneity and stochasticity inherent to the intrinsic randomness of biochemical interactions,
which are forced to take place in a small and crowded bioreactor such as the cell.
A tool frequently used to
face this complexity is mathematical modeling, which allows us to test what possible behaviors arise from
a given molecular network. In the words of the computational cell biologist John Tyson, trying to understand
living systems is ``something like finding a jumble of jigsaw puzzle pieces in a paper bag". We
do not have the model of the picture that we should put together, we are not sure whether all the pieces
are in the bag, and we do not even have a table to test if the pieces fit together. ``Mathematical
modeling provides the table" \cite{Tyson:1999p977}. But the use of models highlights another
problem that arises from the large size of typical gene regulatory networks: the number of parameters to
be adjusted is frequently tremendously large, thus the problem of fitting the model to experimental
data becomes seriously underdetermined.

A second factor that prevents us from relating the architecture of gene regulatory
networks with cellular function is the fact that cellular processes are strongly {\em dynamic}.
Indeed, protein expression in cells varies in
general with time, due either to temporal changes in the external conditions of the cells
(such as circadian rhythms originated in the suprachiasmatic nuclei of mammals, which affect the
rest of cells in the organism via the endocrine system), or to self-generated dynamical behavior
(such as the cell cycle in dividing cells). Thus the complex networks of genes and proteins
mentioned above, and the intricate pattern of interactions among them, are far from being
static. It is therefore necessary to take into account that not all network connections are
active at all times, and that it is the dynamical pattern of connectivity what has to be related
with cellular function.

Fortunately, the dynamic character of gene regulation alleviates the problems caused
by the complexity of the underlying genetic networks. First, nonlinear physics tells us
that dynamical behavior does not require a large number of degrees of freedom. Oscillations,
for instance, can be generated with only two degrees of freedom in continuous dynamical
systems, or even with only one degree of freedom if delays are involved. Therefore, it
is reasonable to expect (from the point of view of biological efficiency) that periodic
cellular behaviors (such as circadian rhythmicity, or cell-cycle
oscillations) are generated
by small core genetic modules embedded in (much) larger genetic networks, with the
additional proteins of the network being used to modulate the dynamics and couple it
with the rest of the cell's physiology. This approach follows the wake of a new
{\em modular} perspective in cell biology, that has been unraveling in the last
decade \cite{Hartwell:1999fk,Tyson:2003uq,Alon:2007p673}. Within this context, one can conjecture
that dynamical cellular functions are governed by core modules of small numbers
of genes and proteins, which we will call {\em genetic circuits} in what follows.
In certain cases, such as in negative feedback loops, a careful analysis
of the dynamical traits of the circuit's behavior may be used to determine the precise
structure of the circuit (i.e. the specific nature of each interaction) \cite{Pigolotti:2007p643}.

A second benefit of the dynamical character of cellular processes regards the issue
of parameter fitting. Interestingly, parameters of dynamical
systems are much more strongly constrained when the system exhibits a time-dependent
state than when it operates in a fixed-point attractor \cite{Kirk:2008kx}. Indeed,
a constant level of protein expression may be generated in multiple
ways, while when protein levels exhibit a certain non-steady dynamics the number
of circuits responsible (and the associated set of parameter values) is much more restricted.

Thus, dynamical phenomena in cells help us determine the molecular mechanisms of
cell function by (i) allowing us to extract, from complex genetic networks, circuits
with small numbers of elements that function as core modules generating
the dynamical response of interest, and (ii) constraining the values of the (usually many)
circuit parameters that better represent the operating state of the cell.
This process requires a close interaction between theory and experiments, where
experimental data is used to build a theoretical model, which is then validated
by testing whether new model predictions are verified experimentally, while
the new experimental data obtained in this process is used to refine the model,
so that the discovery cycle starts again \cite{Kitano:2002p973}.
In this Review, we describe several
case examples where this method has been employed successfully, together
with other phenomenology where it could be used.

Physical approaches to biological problems have a long history. However,
only recently experimental data at the systems level has started to become
available at a scale sufficient to allow for an understanding of how living systems
(self-)organize. Using this experimental information, physicists are currently
applying, for instance, methods from the theory of nonlinear oscillations and bifurcation
analysis to understand the emergence of dynamical phenomena in cells, including periodic
oscillations (such as cell cycle oscillations, circadian rhythms, and many transcription-factor
oscillations) and excitable behavior. Another topic where physics is having a large impact
is the identification of stochasticity in cellular behavior, and the characterization of its
influence. Dynamical processes at the level of gene regulation are
unavoidably affected by considerable levels of random fluctuations, or noise,
which arise from the small number of biomolecules that participate in the
biochemical reactions underlying the cell's physiology.
Within that context, results and approaches from the theory of stochastic
processes, developed during the past four decades by the stochastic physics community,
are starting to be routinely applied to biological problems. The goal is to
interpret experimental observations that reveal
high levels of variability, both in time and space, in the operation of genetic circuits.
Finally, another biological question where physics is having a large impact is in the
study of collective cellular phenomena, in which the theory of coupled oscillators is
playing a key role. When the genetic circuits in individual cells are intrinsically
dynamic, cell-cell coupling provides a mechanism for the spatial
organization of the dynamics. In particular, relevant issues in this respect are the emergence of
synchronization and clustering in cellular populations. These phenomena are well
known in physical systems, and are now revealing themselves as biologically meaningful
dynamical states that drive the spatiotemporal behavior of multicellular systems.

This article presents an overview of the biological phenomena that can be addressed
by the physical approaches discussed in the previous paragraph. The emphasis is
placed more on the biological questions than in the analysis techniques themselves,
since these techniques are rather well established in their application to traditional
physics problems. The review begins by describing
in Sec.~\ref{sec:tools} the most common theoretical and experimental methods that are
being used in those studies, followed in Sec.~\ref{sec:dyn} by examples of
dynamical processes in genetic circuits that have been identified experimentally
in recent years. Section~\ref{sec:noise2} describes the experimental methods that
have been devised in recent years to quantify and characterize stochastic fluctuations
in cells. This Section also discusses how standard theoretical approaches to
stochasticity are being applied to understand the effects of noise in genetic circuits.
Finally, Sec.~\ref{sec:coupling} describes how macroscopic behavior
emerges in cell populations from the organized activity of the individual
genetic circuits operating in each cell. The different types of cell-cell communication
mechanisms that couple these circuits together are discussed, and
recent studies of cell population dynamics and its potential role
in the organization of multicellular organisms are reviewed.

\section{Characterizing dynamical phenomena in genetic circuits}\label{sec:tools}

The behavior of a living cell (its {\em phenotype}) is governed mainly
by the collection of distinct proteins existing within the cell at any given time.
Proteins are the basic components of the majority of structural elements inside the cell, and
perform most of the catalytic biochemical reactions on which life depends.
Each protein is created (expressed) from a given gene via {\em transcription} of that gene into 
a messenger RNA (mRNA) molecule, which is further {\em translated} into the sequence
of amino-acids that constitutes the protein. The entire process of protein expression is subject
to {\em regulation}, specially by proteins themselves. By way of example, certain
proteins called {\em transcription factors} bind to small pieces of DNA
known as {\em promoters}, which precede a given gene and enhance its expression
by promoting the recruitment of the enzyme RNA polymerase (RNAp), which
transcribes the gene into mRNA. These transcription factors are thus called
{\em activators}. A second class of transcription factors ({\em repressors}) 
interfere with RNAp binding and thus inhibit transcription.

\setlength{\fboxsep}{2mm}
\begin{center}
\colorbox{grey}{\parbox{0.92\linewidth}{\small
\textsc{Glossary: terms commonly used in the study of genetic circuits}

\vskip1mm
\textbf{\textit{Antibody --}} protein used by the immune system to identify foreign molecules, in particular other
proteins, by binding to them with high specificity. Antibodies are routinely used, in combination with different
kinds of markers, to monitor the presence of proteins of interest in a sample via multiple methods, as
discussed in Sec.~\ref{sec:exp}.

\vskip1mm
\textbf{\textit{Apoptosis --}} molecular process through which a cell kills itself.

\vskip1mm
\textbf{\textit{Enzyme --}} protein that increases the rate of occurrence of (i.e. catalyzes) a given chemical reaction.
The input of the reaction is called the substrate, and the output is called the product.

\vskip1mm
\textbf{\textit{Eukaryotic cell --}} cell in which the genetic material is enclosed in a nucleus (e.g. yeast and the cells of higher organisms, including animals and plants).

\vskip1mm
\textbf{\textit{Gel electrophoresis --}} technique used for the separation of biological macromolecules such as DNA,
RNA or proteins, in terms of their mobility through an agarose gel under the action of an electric field.

\vskip1mm
\textbf{\textit{Lysis --}} process through which a cellular membrane is broken open, so that the inner
components of the cell are spilt to the extracellular medium.

\vskip1mm
\textbf{\textit{Pluripotency --}} property of a stem cell that allows it to differentiate into any cell type of the
body. It should be distinguished from \textbf{\textit{totipotency}}, which arises when the cell can also generate
extraembryonic tissue (such as the placenta), and from \textbf{\textit{multipotency}}, which refers to the potential
to generate several cell types within a limited number of lineages (such as different types of blood cells,
which arise from multipotent hematopoietic stem cells).

\vskip1mm
\textbf{\textit{Polymerase chain reaction (PCR) --}} iterative process through which the number
of copies of a DNA fragment is increased exponentially. It
is based on the separation of the two DNA strands (denaturation), followed by the synthesis of the
corresponding complementary strands (so that one copy of the DNA fragment produces two),
after which the process is repeated again giving rise to a chain reaction.

\vskip1mm
\textbf{\textit{Prokaryotic cell --}} cell without nucleus or other membrane-bound organelles (mainly bacteria).

\vskip1mm
\textbf{\textit{Promoter --}} piece of DNA usually preceding a gene, to which RNA
polymerase and transcription factors bind, thus
controling expression of the gene.

\vskip1mm
\textbf{\textit{Ribosome --}} cellular complex that assembles proteins using an mRNA molecule as a template.
The ribosome itself is formed by both proteins and RNA.

\vskip1mm
\textbf{\textit{RNA polymerase --}} enzyme that assembles RNA molecules using a DNA strand as a template.

\vskip1mm
\textbf{\textit{Suprachiasmatic nucleus --}} collection of around 10,000 neurons located right above
the optical chiasm in the brain, responsible for the generation of circadian rhythms in
higher organisms.

\vskip1mm
\textbf{\textit{Transcription --}} process through which an mRNA molecule is created from a gene.
 
\vskip1mm
\textbf{\textit{Transcription factor --}} protein that controls the transcription of a given gene by binding to
its promoter, either activating or inhibiting it.

\vskip1mm
\textbf{\textit{Translation --}} process through which a protein is generated from an mRNA molecule.
}}
\end{center}

Regulation can also
arise at the post-transcriptional level, for instance by means of {\em RNA interference},
which arises when small RNA molecules attach to a
given mRNA molecule and induce its degradation \cite{Mcmanus:2002p1367}.
Finally, regulation takes place at well at
the post-translational level, through processes such as phosphorylation (which
usually turns a protein into its active form) and proteolytic degradation.

In any case, gene regulation provides a rich
source of feedback in the expression process, either at the level of a single
protein (i.e. of a protein on itself) or at the network level (i.e. of a protein on
others that might, in turn, act back upon it). Considering that
expression and regulation processes are frequently nonlinear (see Sec.~\ref{sec:models}
below), these feedbacks
lead to a rich amount of complexity, which allows for instance the coexistence of multiple
phenotypes within the same {\em genotype} (think for instance of the many distinct cell types
in our body).
In this Section, we present a brief overview of the methods, both theoretical and
experimental, used to describe the dynamical behavior of small collections of genes
and proteins interacting with one another via gene regulation processes.

\subsection{Theoretical tools}\label{sec:models}

\subsubsection{Deterministic description of gene-circuit dynamics}
\label{sec:lma}

The expression and regulation processes described in the previous paragraphs can be
represented mathematically, in a first approximation, by means of standard rules
of reaction kinetics. These rules are grounded on the {\em law of mass action}, according
to which the rate of a chemical elementary process is proportional to the product of
the concentrations of the molecular species involved in the process, the proportionality
constant corresponding to the reaction rate.  
Consider for instance the well-known case of the simple
enzyme catalysis reaction:
$$
E+S\;\autorightleftharpoons{$k_{1}$}{$k_{-1}$}\;[ES]\;\autorightarrow{$k_{\rm tr}$}{}\;E+P
$$
Here $E$ represents the enzyme, which transforms a substrate $S$ into a product $P$
through the creation of a complex $[ES]$.
Interestingly, this reaction scheme can also represent transcriptional regulation in
its simplest form,
provided $E$ is interpreted as the gene promoter, $S$ as the transcription factor, and $P$
as the mRNA that results from the transcription process\footnote{in this case $S$ would also
be a product of the reaction producing $P$, since the transcription factor does
not disappear from the system upon transcription, but this will not affect the discussion
below.}. We will refer to this latter interpretation, without loss of generality, in what follows.

\begin{figure}[htb]
\centerline{\includegraphics[width=0.4\textwidth]{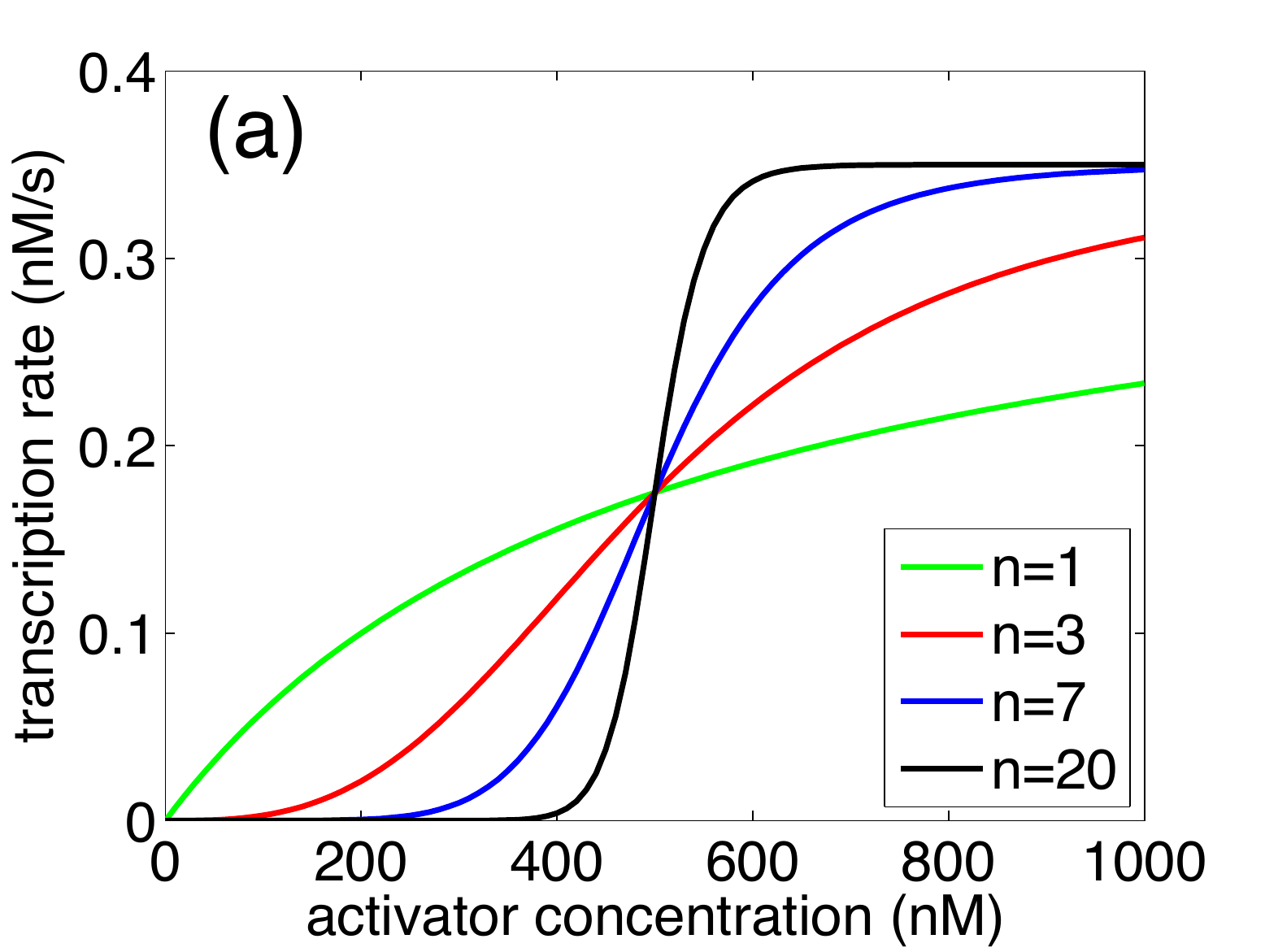}
\includegraphics[width=0.4\textwidth]{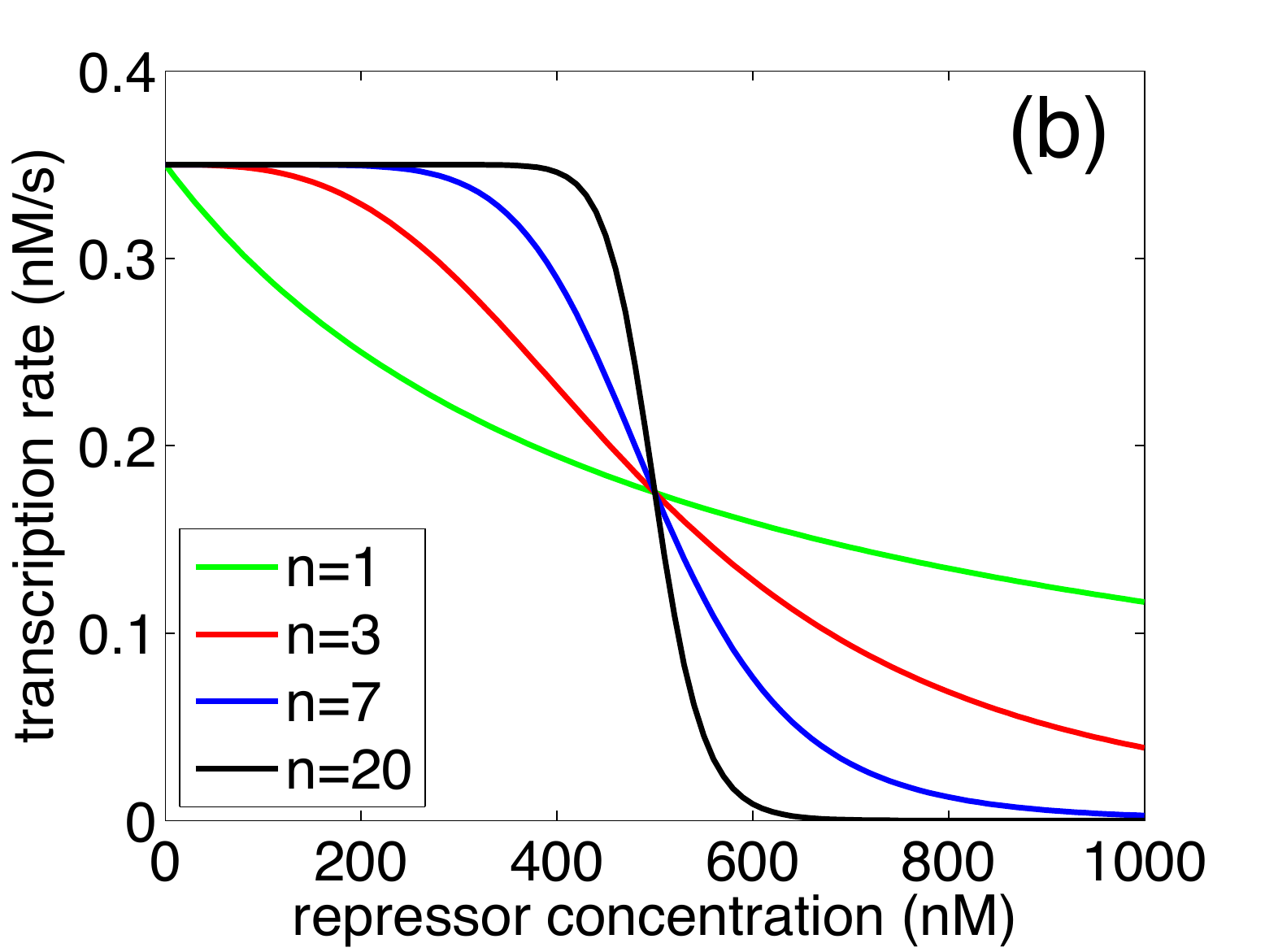}}
\caption{Rate of mRNA production (transcription rate, $dP/dt$) versus transcription
factor concentration $S$ for the cases of activation [plot (a), Eq. (\ref{eq:Pactcoop})]
and inhibition [plot (b), Eq. (\ref{eq:Prep})]. Different lines correspond to different
cooperativity coefficients $n$. Parameter values are $\beta=0.35$~nM/s,
$k_{\rm act}=k_{\rm rep}=500$~nM.
}
\label{fig:hill}
\end{figure}

The equation describing the dynamics of the complex formed by the
promoter and the transcription factor 
can be derived from reaction kinetics:
\begin{equation}
\frac{d[ES]}{dt}=k_1E\cdot S-(k_{-1}+k_{\rm tr})[ES]\,,
\label{eq:ES}
\end{equation}
where now $E$, $S$, and $[ES]$ represent the concentrations of the corresponding
species. Assuming that the complex relaxes to its instantaneous steady state ($d[ES]/dt=0$)
much faster than the other species, and taking into account that the promoter
copy number does not change during the transcription process ($E+[ES]=E_T$),
one can establish the dependence of the instantaneous concentration
of the complex in terms of the transcription-factor concentration as
\begin{equation}
\frac{d[ES]}{dt}=0 \quad\Longrightarrow\quad
[ES]=\frac{ k_1E_TS}{k_{-1}+k_{\rm tr}+k_1S}=\frac{E_TS}{k_{\rm act}+S}\,,
\label{eq:ESss}
\end{equation}
where $k_{\rm act}=(k_{-1}+k_{\rm tr})/k_1$. This is referred to as the {\em adiabatic
elimination} of the fast variable $[ES]$. With this approximation, the rate of mRNA production
can be calculated as
\begin{equation}
\frac{dP}{dt}=k_{\rm tr}[ES]=\frac{\beta S}{k_{\rm act}+S}\,,
\label{eq:Pact}
\end{equation}
where $\beta=k_{\rm tr}E_T$ depends both on the rate of production of
mRNA from the complex and on the promoter copy number\footnote{In the context of
enzyme kinetics, this equation is known as the {\em Michaelis-Menten equation}.}. When
multiple units of the transcription factor $S$ are necessary to activate transcription
(a condition known as {\em cooperativity}),
and assuming that its multimerization (taking place either prior to or after binding to the
promoter) occurs again much faster than other reactions in the process, one can
reach in a straightforward way the following expression for the mRNA production rate:
\begin{equation}
\frac{dP}{dt}=\frac{\beta S^n}{k_{\rm act}^n+S^n}\,.
\label{eq:Pactcoop}
\end{equation}
This expression corresponds to a Hill function, with the Hill coefficient $n$
representing the degree of cooperativity of the transcription process.
Figure~\ref{fig:hill}(a) represents the mRNA production rate $dP/dt$ (which
we will also call {\em promoter activity} in what follows) versus the transcription
factor concentration $S$ for increasing levels of cooperativity $n$. The figure clearly
shows that a high enough cooperativity leads to an abrupt transcriptional
switch, with $k_{\rm act}=[(k_{-1}+k_{\rm tr})/k_1]^{1/n}$ being the activation threshold.

The case of transcriptional repression can be treated analogously, with the
basic reaction scheme and resulting mRNA production rate being
\begin{equation}
E+nS\;\autorightleftharpoons{$k_{1}$}{$k_{-1}$}\;[ES^n];
\quad
E\;\autorightarrow{$k_{\rm tr}$}{}\;E+P
\qquad\Longrightarrow\qquad
\frac{dP}{dt}=\frac{\beta}{1+(S/k_{\rm rep})^n}\,,
\label{eq:Prep}
\end{equation}
where the repression threshold is $k_{\rm rep}=(k_{-1}/k_1)^{1/n}$.
Figure~\ref{fig:hill}(b) shows the production rate of mRNA for the case of
transcriptional repressor, for increasing concentrations of repressor and
for varying levels of cooperativity. The switch-like character of the repression
for high enough cooperativity is again obvious.

The expressions given in Eqs.~(\ref{eq:Pact}) and (\ref{eq:Prep})
describe the creation of mRNA molecules, which will be later
translated into proteins by ribosomes. In general we can assume the
protein production rate to be proportional to the mRNA concentration,
provided we neglect translational regulation. On the other hand, dilution of
the mRNA and protein molecules due to cell growth (and eventually
cell division) lead to a concentration decrease at a rate that can be considered
proportional to the concentration, barring any regulated degradation
mechanism. 

\paragraph{One-dimensional dynamics}

\begin{figure}[htb]
\centerline{\includegraphics[width=0.67\textwidth]{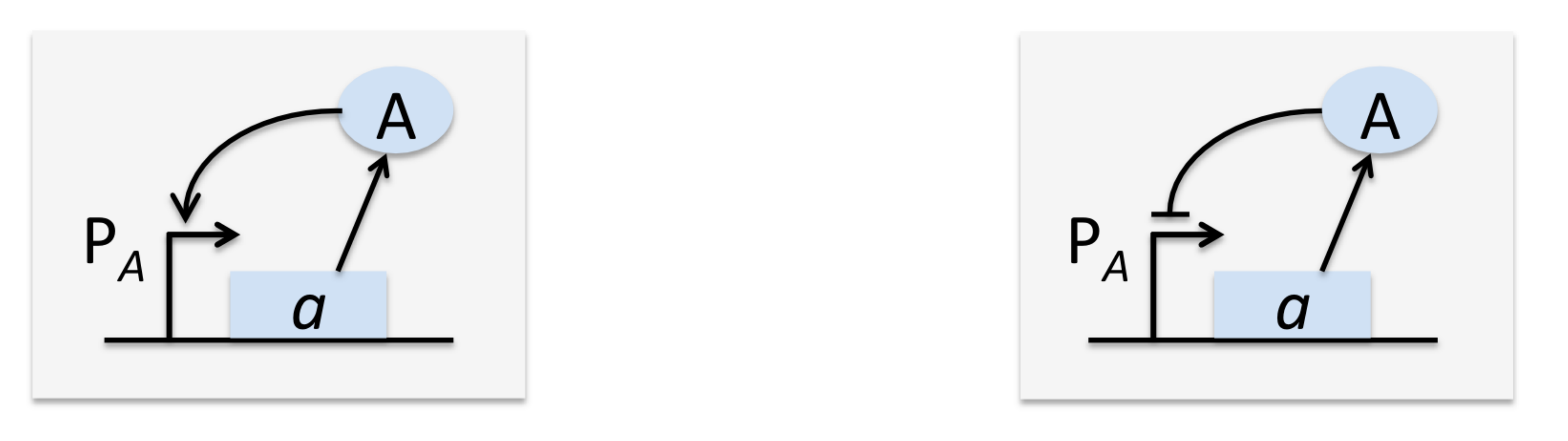}}
\centerline{\includegraphics[width=0.45\textwidth]{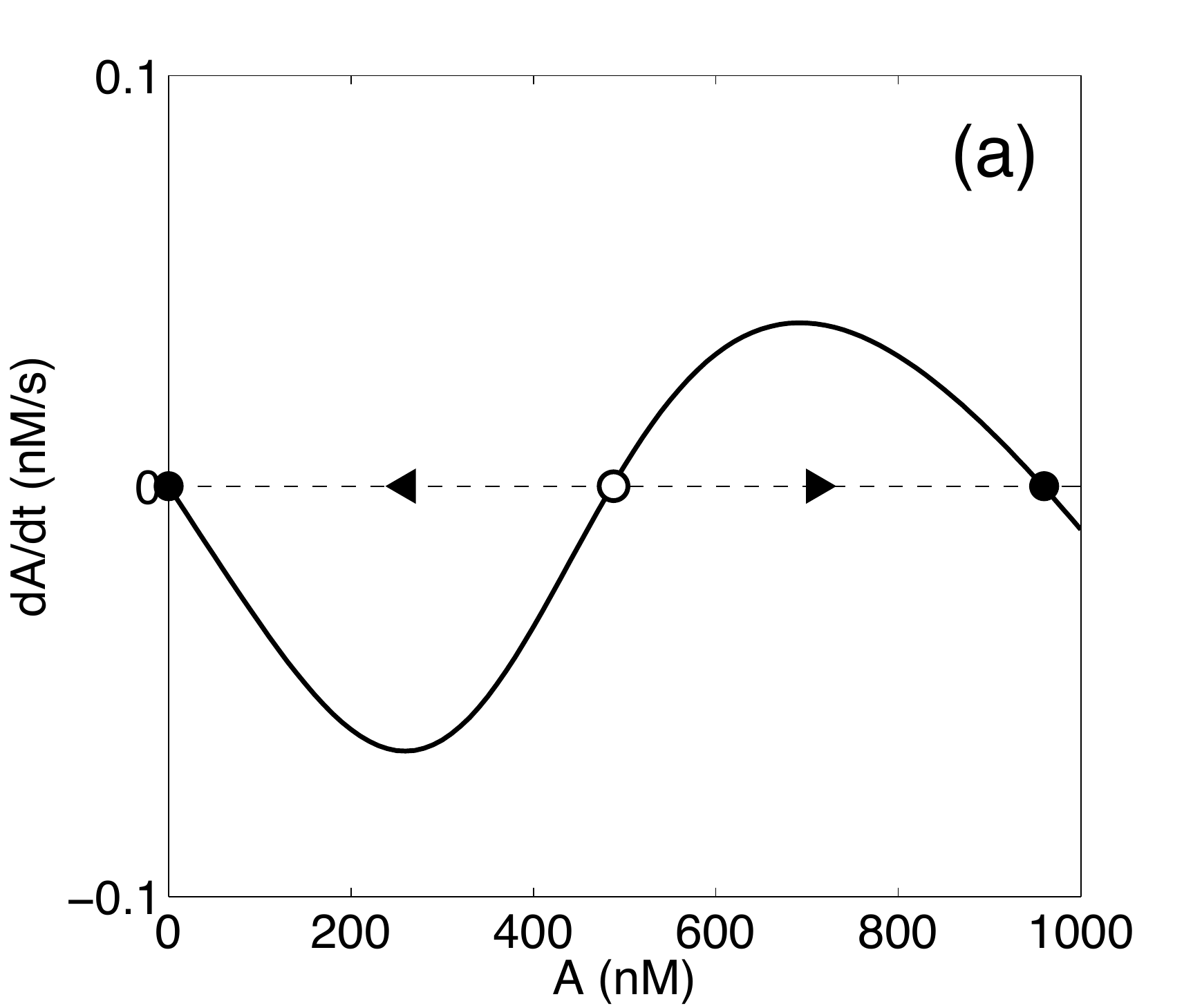}
\includegraphics[width=0.45\textwidth]{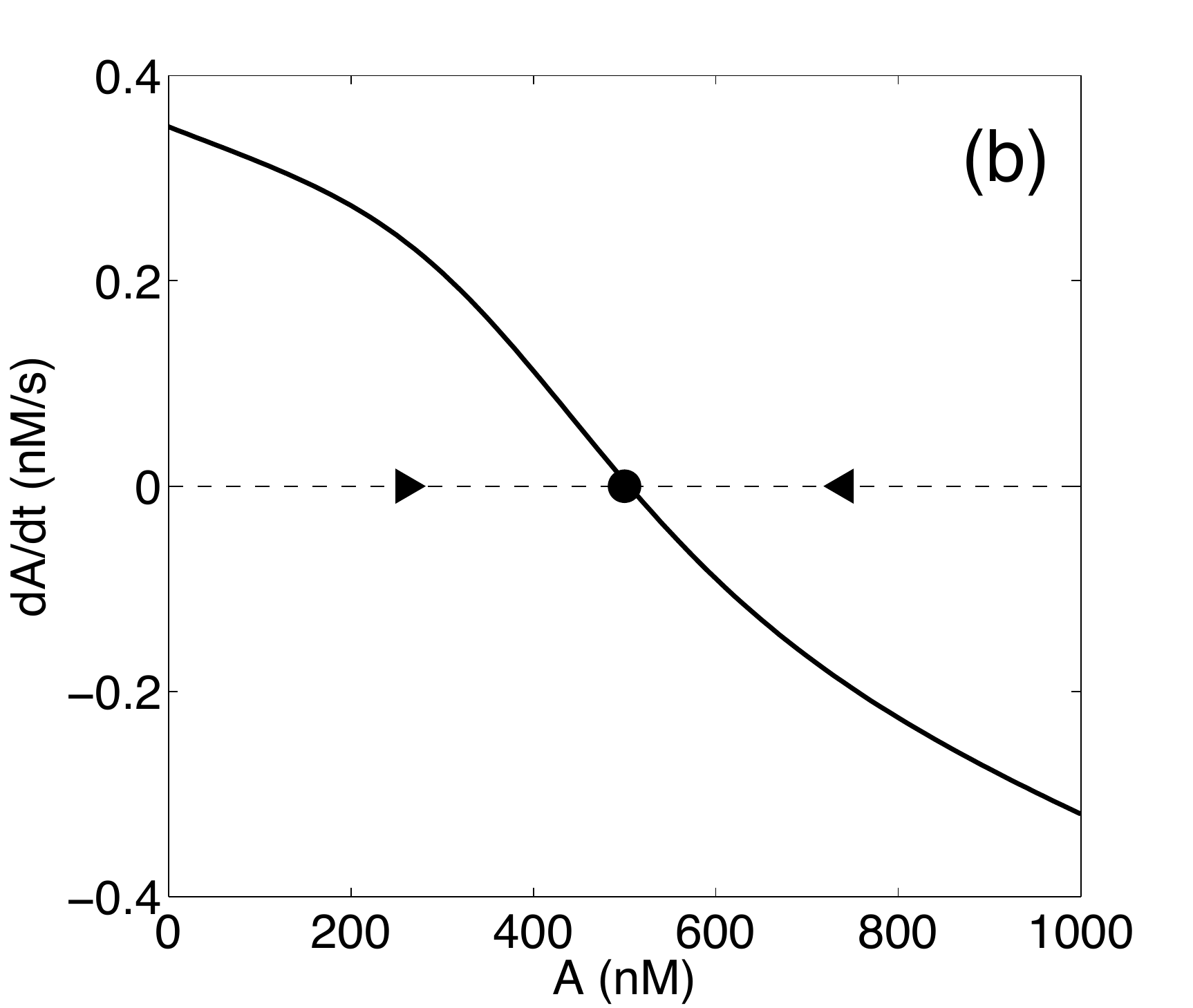}}
\caption{Phase-line diagrams of
the dynamical equations of a self-activating (a) and a self-repressing (b) circuit.
Solid (empty) circles correspond to stable (unstable) fixed points.
The two cartoons in the top row show schemes of the corresponding genetic circuits. In those
schemes, arrows represent activation and blunt lines denote inhibition.
Parameter values are those of Fig.~\ref{fig:hill} with $n=4$, plus $\delta=1.22$~h$^{-1}$.
}
\label{fig:1d}
\end{figure}

With the previous ingredients we can now model the two simplest genetic
circuits, namely a single protein that either activates or represses its own
expression. Representing by $A$ the protein concentration and by $a$
the corresponding mRNA concentration, we can write the following
dynamical equations for the self-activation circuit [plot on top of Fig.~\ref{fig:1d}(a)]:
\begin{equation}
\frac{da}{dt}=\frac{\beta_{\rm tx} A^n}{k_{\rm act}^n+A^n}-\delta_a a\,,
\qquad
\frac{dA}{dt}=\beta_{\rm tl}a-\delta_A A\,,
\label{eq:selfact}
\end{equation}
where $\beta_{\rm tx}$ is the transcription rate at saturation, $\beta_{\rm tl}$
is the translation coefficient, and $\delta_a$ and $\delta_A$ are the
linear degradation rates of the two species. In bacteria such as {\em E. coli},
mRNA degrades on the order of a few minutes, whereas proteins degrade on
the order of hours (limited by the growth rate) \cite{Alon:2006fk}. In those conditions, we can
assume $a$ to relax quickly to its instantaneous steady state
($da/dt=0$), so that we are left with only one degree of freedom, $A$, whose
dynamics is given by
\begin{equation}
\frac{dA}{dt}=\frac{\beta A^n}{k_{\rm act}^n+A^n}-\delta A\,.
\label{eq:selfact2}
\end{equation}
Here $\beta=\beta_{\rm tx} \beta_{\rm tl}/\delta_a$, and the subindex in the
degradation rate of $A$ has been dropped for compactness. The dynamical
behavior of this circuit can be qualitatively established by plotting the right-hand-side
of Eq.~(\ref{eq:selfact2}) versus the phase line defined by the $A$ axis, as shown in Fig.~\ref{fig:1d}(a).
The zeros of that function correspond to the fixed points of the system, while in
the non-zero regions its sign determines in an unambiguous manner (since
the system is deterministic) the tendency in time (towards growth or
decay\footnote{Note then that a genetic circuit with only one degree of freedom cannot oscillate
(unless there are delays \cite{Bratsun:2005fk}).}) of $A$.
In that way, the sign of the slope of $dA/dt$ at the fixed point dictates the stability
of the fixed point: a negative slope corresponds to a stable fixed point (solid circle
in the figure), towards which the circuit tends at long times. The unstable fixed point
(empty circle) separates in the phase line basins of attraction of different stable states.
The figure clearly shows that for the conditions considered (basically high enough
cooperativity) transcriptional self-activation leads to a bistable response.

The case of self-repression [plot on top of Fig.~\ref{fig:1d}(b)], on the other hand, is represented
by the following
dynamical equation
\begin{equation}
\frac{dA}{dt}=\frac{\beta}{1+(A/k_{\rm rep})^n}-\delta A\,,
\label{eq:selfrep2}
\end{equation}
whose phase-line representation is shown in Fig.~\ref{fig:1d}(b). Only a single stable
steady state exists, corresponding to the homeostatic behavior characteristic of negative
feedback circuits. Multiple examples of these two types of minimal circuits (positive
and negative self-feedbacks) exist in the literature (see for
instance \cite{Alon:2007p673} for a review).

\paragraph{Two-dimensional dynamics}\label{sec:2d}

\begin{figure}
\centerline{\resizebox*{6cm}{!}{\includegraphics[clip]{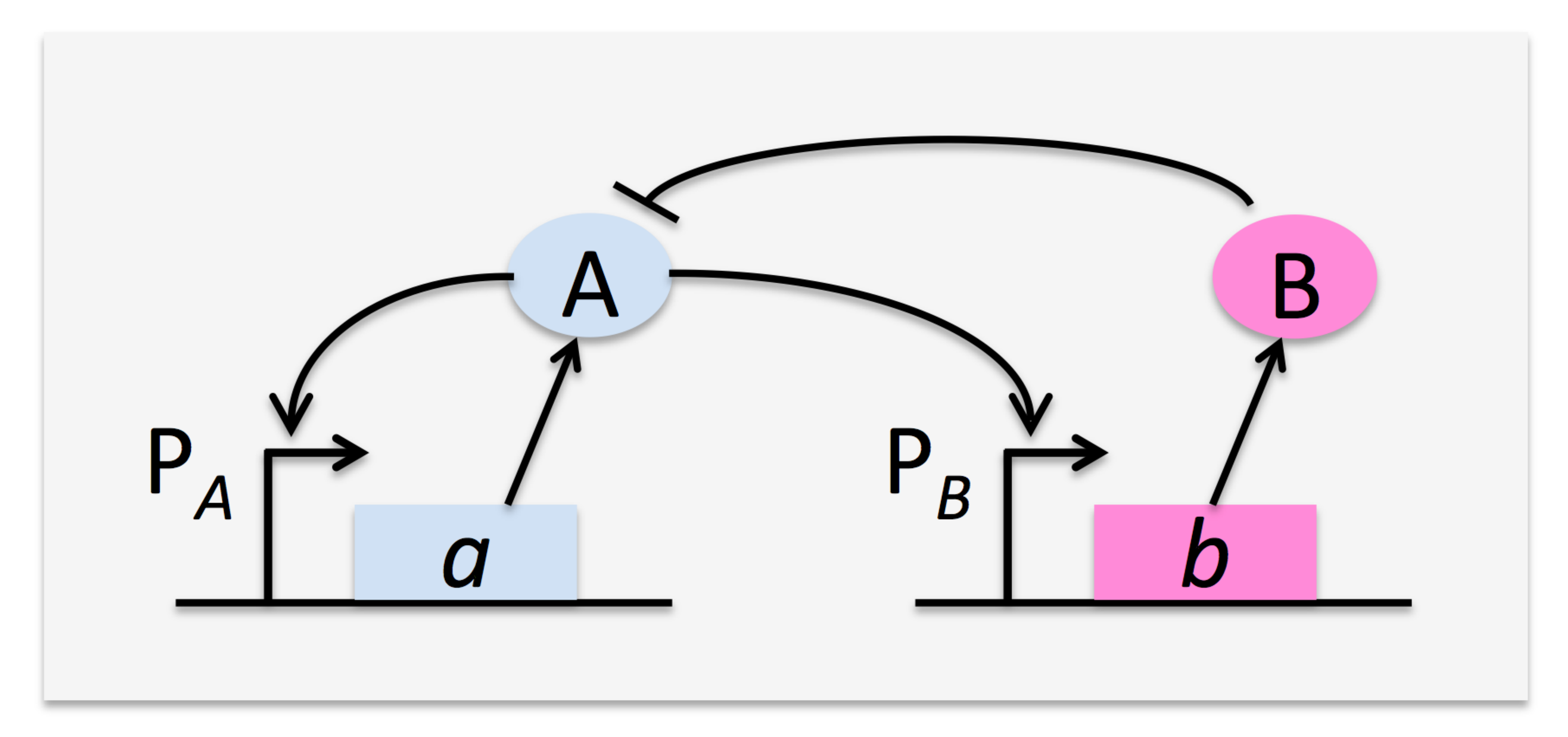}}}
\vskip3mm
\centerline{\resizebox*{9.8cm}{!}{\includegraphics{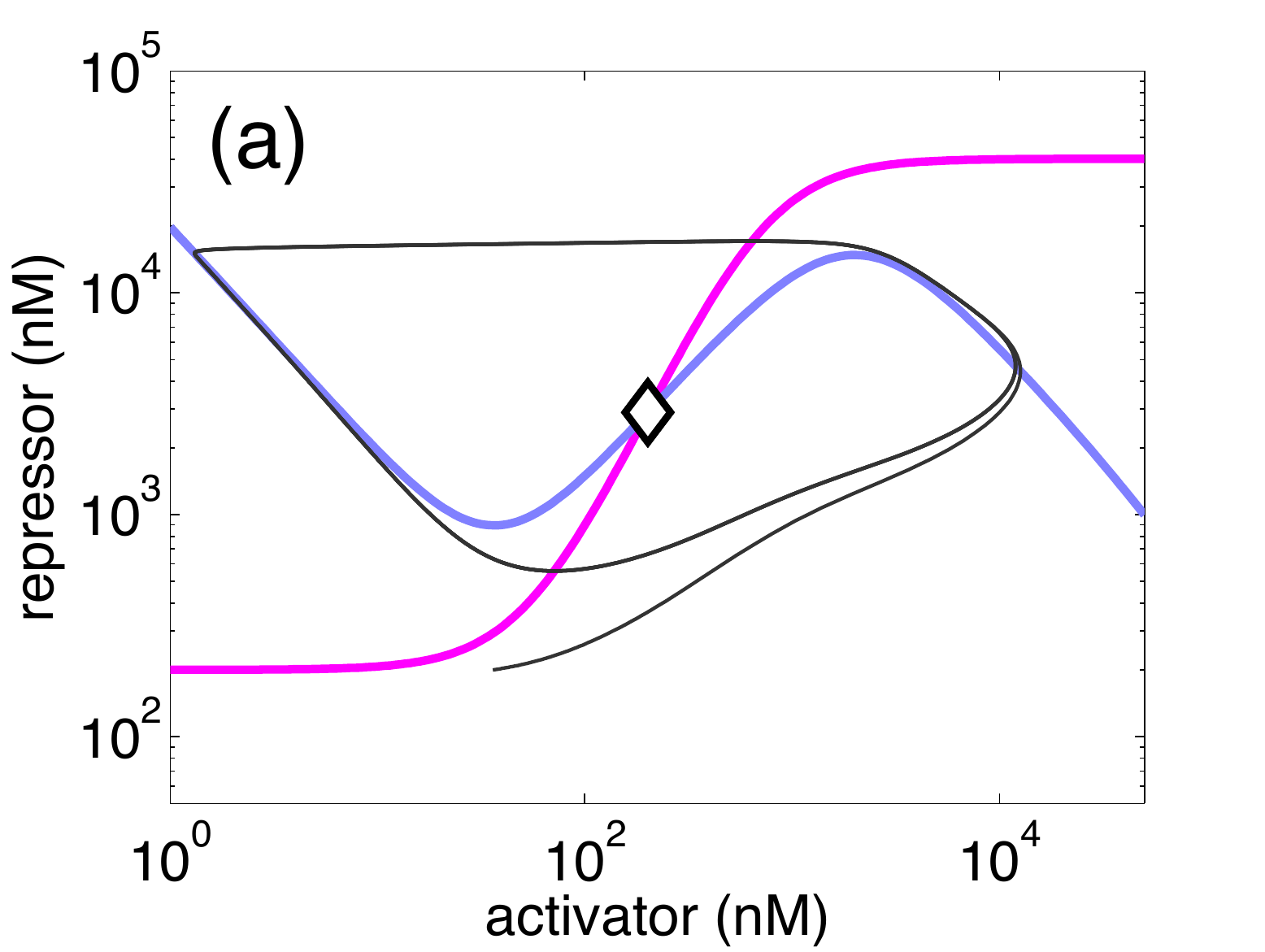}
\includegraphics{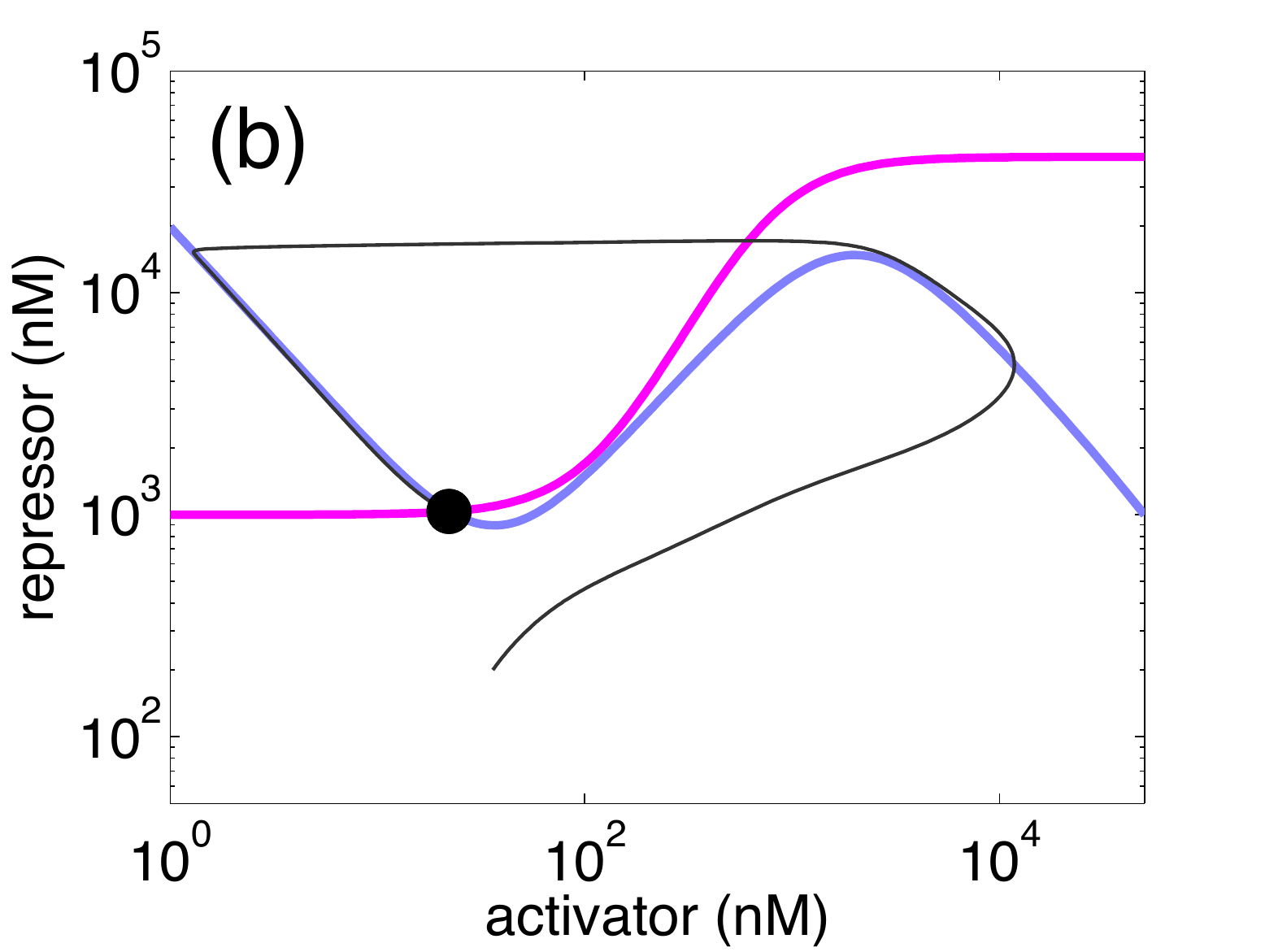}}}
\centerline{\resizebox*{10cm}{!}{\includegraphics{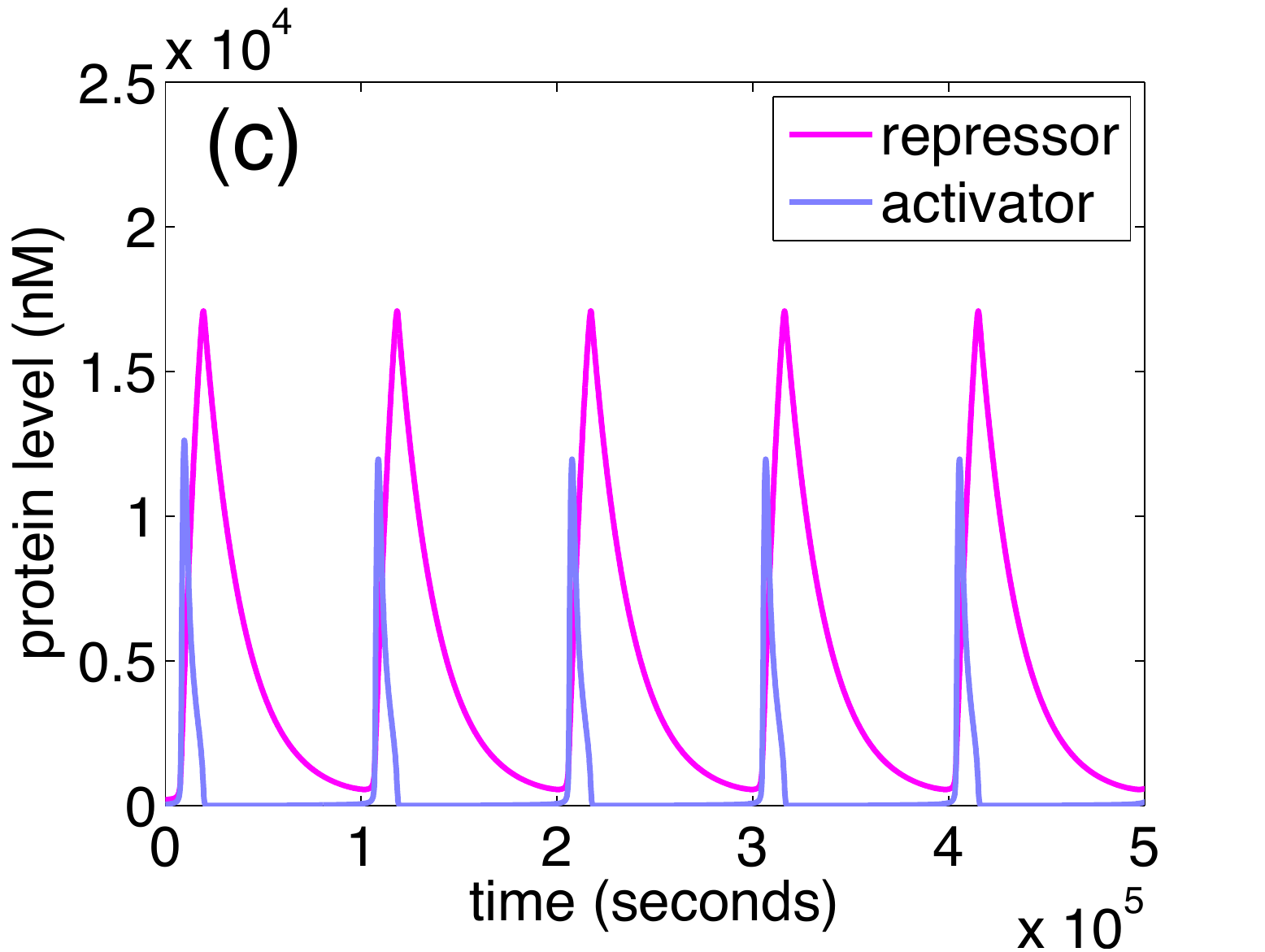}
\includegraphics{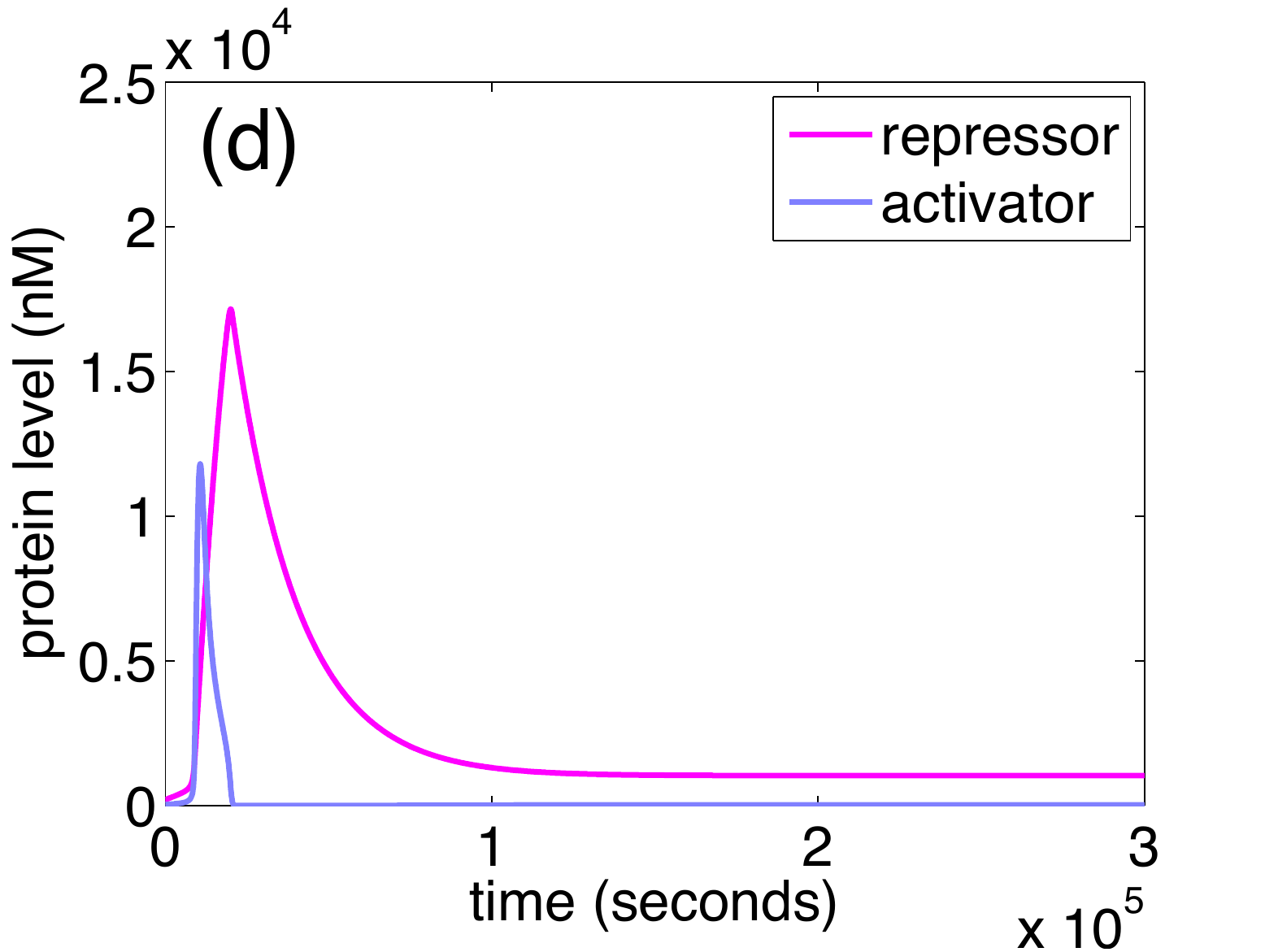}}}
\centerline{\resizebox*{10cm}{!}{\includegraphics{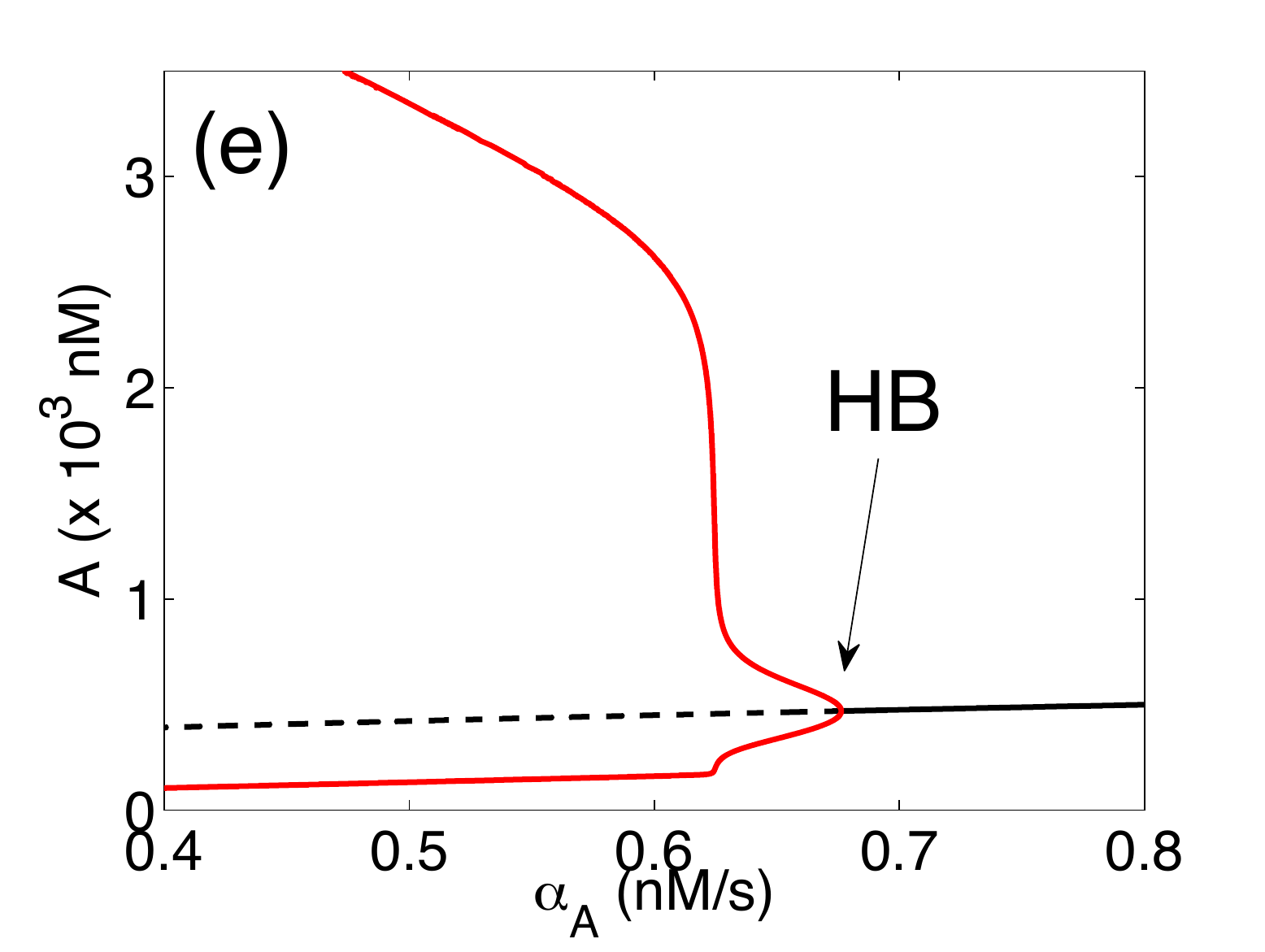}
\includegraphics{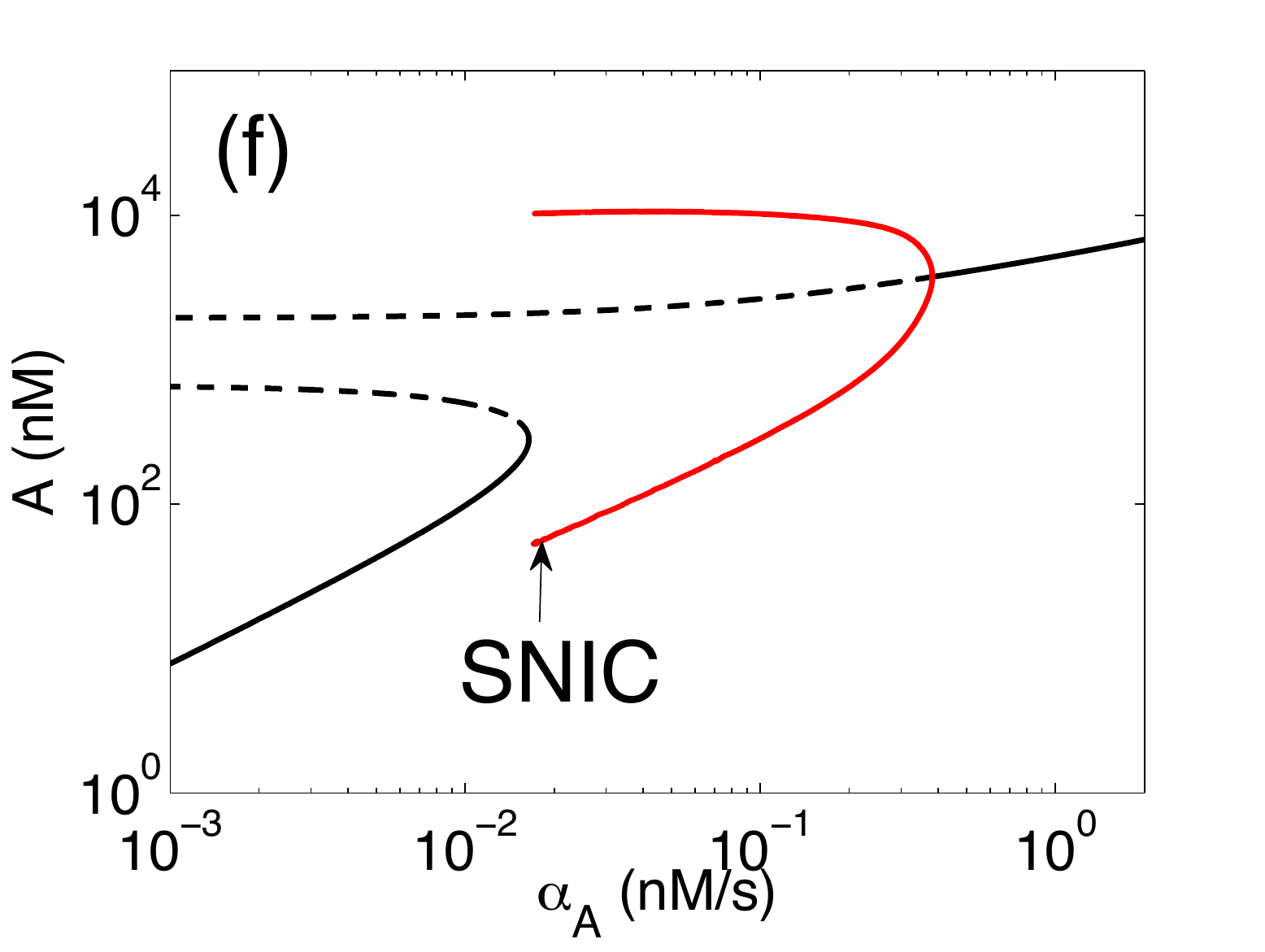}}}
\caption{Dynamics of an activator-repressor genetic circuit (top panel).
Phase-plane portraits for oscillatory (a) and excitable (b) dynamics are
shown. In those plots, the magenta and pale blue lines correspond to the nullclines,
the thin black line to a sample trajectory, the empty diamond to an unstable
spiral point, and the solid circle to a stable fixed point. Plots (c) and (d)
show time traces corresponding to the two regimes, oscillatory and excitable,
respectively. Finally, plots (e) and (f) represent two different bifurcation
routes leading to genetic oscillations. In those plots, the solid black line
denotes a branch of stable fixed points, the dashed one corresponds to
unstable fixed points, and the red line stands for a branch of stable limit
cycles.
The common parameter values are $\alpha_A =  0.005$~nM/s,
$\alpha_B = 0.01$~nM/s, $\beta_A =  15$~nM/s,
$\beta_B = 2$~nM/s, $g = 2.5\cdot 10^{-7}$~nM$^{-1}$s$^{-1}$,
$\delta_A = 5\cdot 10^{-5}$~s$^{-1}$, $\delta_B = 5\cdot 10^{-5}$~s$^{-1}$,
$k_A = 2000$~nM, $k_B = 750$~nM, $n = p= 2$.
}
\label{fig:2d}
\end{figure}

Including only one additional species in a genetic circuit leads to a qualitative increase
in the richness of the system's behavior. The simplest example is given by the
activator-repressor system depicted in the top panel of Fig.~\ref{fig:2d}: a protein
A activates its own expression and that of a second protein B, which in turn
inhibits the activity of protein A. There are different ways in which such a circuit
can be implemented, depending in particular on whether the different interactions
are transcriptional or post-transcriptional. We now consider one of the most
common situations, in which both activations occur at the level of transcription,
while the repression of A by B takes place at the protein level, for instance by
means of the degradation of A being induced by B.

Using the rules described in the previous Section, which include the adiabatic
elimination of the dynamics of the mRNA of A and B, the dynamical equations
that describe the behavior of the activator-repressor circuit of Fig.~\ref{fig:2d}
are:
\begin{eqnarray}
&&\frac{dA}{dt} = \alpha_A+\frac{\beta_A A^n}{k_A^n+A^n}-gAB-\delta_AA\\
&&\frac{dB}{dt} = \alpha_B+\frac{\beta_B A^p}{k_B^p+A^p}-\delta_BB
\label{eq:actrep}
\end{eqnarray}
Here we have considered that the regulated promoters from which A and B
are expressed are both {\em leaky}, since they can exhibit a certain level
of activity (which we assume below to be relatively small), measured
by $\alpha_A$ and $\alpha_B$, even in the absence of the activator A.

In order to determine the type of behavior exhibited by this system,
we can represent, in the phase plane formed by the protein concentrations
$A$ and $B$, the loci of points for which the respective time derivatives are
equal to zero, known as {\em nullclines}. These curves separate regions in phase
space with differing tendencies of growth/decay of the two proteins (depending
on the sign of their derivatives), and their crossings correspond to
fixed points whose stability depends on the signs of the eigenvalues of the Jacobian
matrix at those points \cite{Strogatz:1994fk}. Figures~\ref{fig:2d}(a-d) display
two qualitatively different behaviors that the activator-repressor circuit
can exhibit. In Fig.~\ref{fig:2d}(a), the nullclines cross at a fixed point which happens to be
unstable, and the circuit is eventually forced to oscillate, with the repressor chasing
(and silencing) activator pulses in a periodic manner. Figure~\ref{fig:2d}(c) shows the
time series corresponding to this {\em limit cycle} behavior. 

In the regime depicted in Figs.~\ref{fig:2d}(b,d),
on the other hand, the nullclines cross at a stable fixed point, but the relaxation
towards this stable state takes the form, for perturbations above a certain threshold, of a
large excursion in phase space. This so-called {\em excitable} regime exists
in a region of parameter space close to the oscillatory regime. There are several
ways in which the oscillations can emerge in a two-component genetic
circuit such as the one described above. This can be revealed by making use
of numerical continuation software, such as AUTO \cite{Doedel:1984fk}, which
allows us to track how the circuit's invariant sets
(either fixed points or periodic orbits) change as a given parameter (or set of parameters)
varies. Figures~\ref{fig:2d}(e,f) show two
{\em bifurcation diagrams} exemplifying two different scenarios for the emergence of
oscillations and excitability. In Fig.~\ref{fig:2d}(e) the limit cycle emerges from a
supercritical Hopf bifurcation (HB) as the constitutive expression rate of A, $\alpha_A$,
decreases. In that case, the limit cycle (red branch) is born with zero amplitude (and non-zero
frequency, see for instance \cite{Strogatz:1994fk}). Excitable dynamics exists
at the right of the bifurcation and close to it.

In the scenario of Fig.~\ref{fig:2d}(f),
on the other hand, the oscillations are born from a saddle-node on an invariant circle 
(SNIC) bifurcation as $\alpha_A$ increases from the left, and they die via a
supercritical Hopf bifurcation as that parameter increases further. In the SNIC bifurcation,
a standard saddle-node bifurcation, in which a stable node and a saddle
mutually annihilate after colliding, takes place directly on top of a limit cycle, which
disappears (or is created, depending on the direction on which the bifurcation
is traversed) as a result of the collision. The limit cycle is born, as $\alpha_A$ increases,
with non-zero amplitude (and zero frequency, see for instance \cite{Izhikevich:2006uq}).
Here the excitable dynamics occurs at the
left of the SNIC bifurcation point.

Bifurcation analyses such as the ones described in the preceding paragraphs are powerful
tools to validate mathematical models of genetic circuits, since they provide predictions,
frequently non-trivial, that can be validated experimentally. To that end, an appropriate
bifurcation parameter has to be chosen, which can be perturbed experimentally in
a controlled way. We discuss an example of this approach in Sec.~\ref{sec:exc} below.

Both the dynamical behaviors and the circuit architectures discussed so far in this section
are the basic building blocks of natural genetic circuits, as will be shown in Sec.~\ref{sec:dyn}.
Even though graphical methods such as the ones exemplified in Figs.~\ref{fig:1d} and
\ref{fig:2d} cannot be used when the circuits have more than two degrees of freedom, the
fixed points and their stability can be computed, and numerical continuation methods
can be used to determine the bifurcation structure in which a given dynamical phenomenon
is embedded. In that way, we can use dynamics to determine molecular mechanisms
of cellular function.

\subsubsection{Stochastic description of gene-circuit dynamics}\label{sec:noise}

The description that has been advocated in the previous section, based on
ordinary differential equations, assumes that the number of molecules in the different
species of the circuit is large, so that it can be considered to vary in a
continuous manner as time evolves. Additionally, the cell's contents are presumed to
be well mixed, so that the rules of reaction kinetics can be applied. These assumptions,
however, are frequently implausible. To begin with, one of the species involved in 
the transcription process, namely the DNA promoter, often occurs only in {\em a single copy}
inside the cell. It is thus unrealistic to assume that the transcription reaction can be
described in a continuous manner, unless the number of mRNA molecules produced
per unit time is large.

A frequently used tool to describe the discrete evolution of molecule numbers in genetic
circuits is the direct numerical simulation of the underlying reactions via Monte Carlo
methods \cite{Wilkinson:2009p749}. A highly popular version of this approach was developed in the late
1970s by D. Gillespie \cite{Gillespie:1977p456}, and is still commonly used. In its simplest
implementation, the method relies on iteratively estimating the next biochemical reaction that should take
place, based on the {\em propensities} of all the reactions that form the circuit, and
update all the molecular species accordingly. Many alternative, faster and more efficient,
implementations of Gillespie's method (both exact and approximate) have been developed
over the years (see for instance \cite{Gibson:2000p452,Gillespie:2001p454,Anderson:2007p833}
for the most popular ones). Software packages have been created to perform discrete simulations
using this type of algorithm in a user-friendly manner in different platforms, such as
BioNetS \cite{Adalsteinsson:2004p1236} and Dizzy \cite{Ramsey:2005fk}.
Finally, it is worth mentioning that from this fully discrete description of the circuit dynamic,
one can develop a continuous description, which still incorporates stochastic effects,
in the form of a chemical Langevin equation \cite{Gillespie:2000p453}. From this type
of mesoscopic description, standard analysis tools from statistical mechanics can be used.
For a generic case of $N$ molecular species involved in $M$ biochemical reactions, the
chemical Langevin equation takes the form
\begin{equation}
\frac{dX_i}{dt}=\sum_{j=1}^M\nu_{ji}a_j({\bf X}(t))+\sum_{j=1}^M\nu_{ji}a_j^{1/2}({\bf X}(t))\xi_j(t)
\label{eq:cle}
\end{equation}
Here $X_i(t)$ denotes the molecule number of species $i$ ($i=1\ldots N$), $\nu_{ji}$ represents how
many molecules of species $i$ are produced in reaction $j$, $a_j({\bf X}(t))$ stands for the
propensity of reaction $j$ (which follows the law of mass action described in Sec.~\ref{sec:lma}),
and $\xi_j(t)$ is a Gaussian white noise uncorrelated between different species. Equation~(\ref{eq:cle})
represents a set of coupled stochastic differential equations with multiplicative noise, which
has to be interpreted according to Ito (since the model describes population dynamics \cite{horsthemke}).
Models within the class of Eq.~(\ref{eq:cle}) have been used for years to study multitude of
physical systems (see, for instance, Refs.~\cite{sanmiguel}).
Early systematic studies of this type of stochastic description applied to genetic circuits can be found,
for instance, in Ref.~\cite{Kepler:2001p1369}.

In order to visualize the effects of stochasticity in gene regulation, let us consider an
extremely simple situation in which a gene is {\em constitutively} transcribed and translated
with constant, unregulated rates, and both the resulting mRNA and protein decay linearly
by dilution, as explained above. In a continuous deterministic description, the system is obeyed
by the trivial linear differential equation $dA/dt = \alpha -\delta A$, so that $A$ tends to
a steady state value $\alpha/\delta$ with characteristic time $\delta^{-1}$. The discrete
behavior of this process will have to take into account the transcription, translation, and degradation
reactions:
\begin{eqnarray}
P_A\;&\autorightarrow{$k_{\rm tx}$}{}&\;P_A + a\nonumber\\
a\;&\autorightarrow{$k_{\rm tl}$}{}&\;a+A\nonumber\\
a\;&\autorightarrow{$k_{\rm deg,a}$}{}&\;\emptyset\nonumber\\
A\;&\autorightarrow{$k_{\rm deg,A}$}{}&\;\emptyset\nonumber
\end{eqnarray}

\begin{figure}[htb]
\centerline{\resizebox*{10cm}{!}{\includegraphics{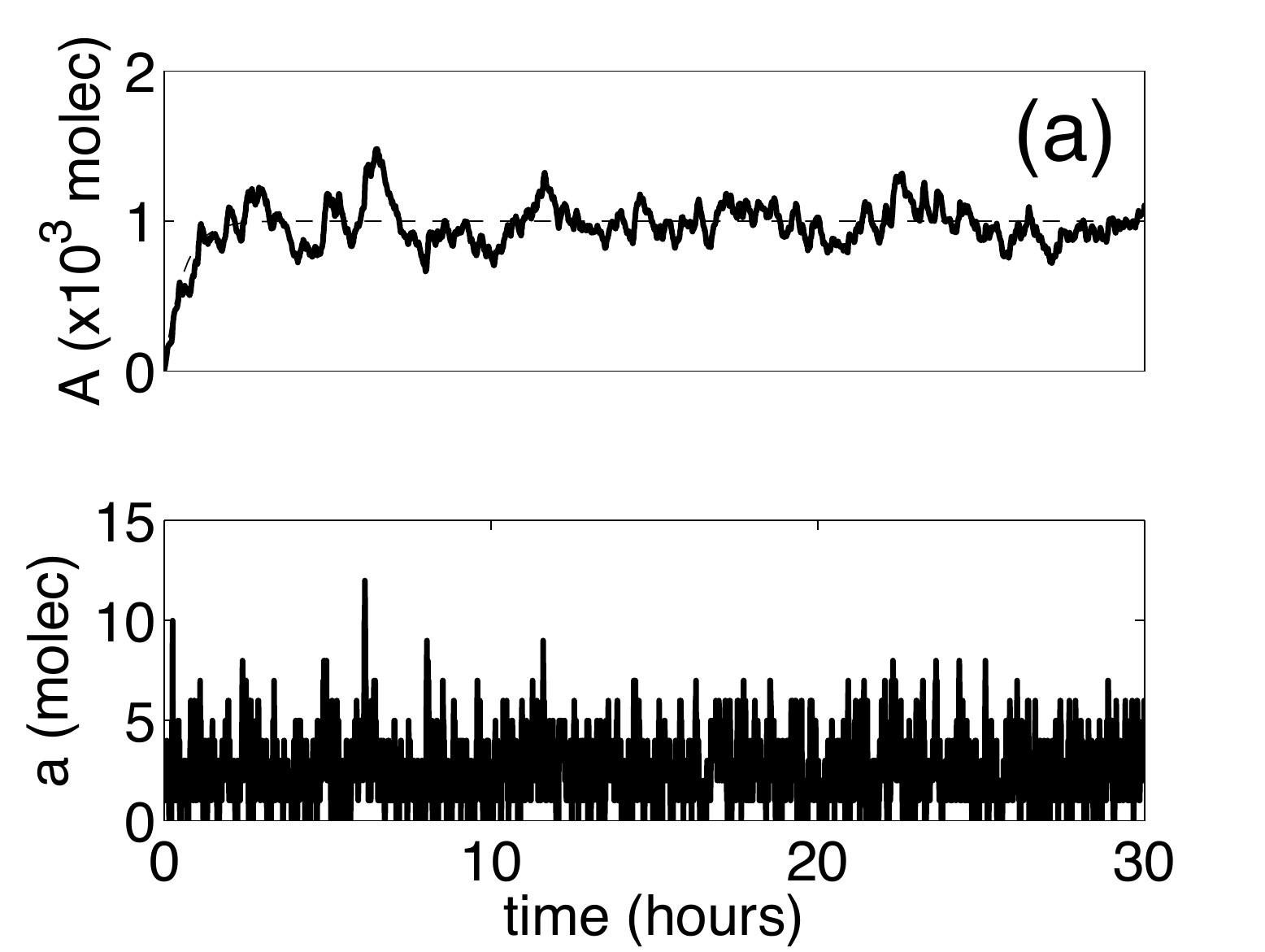}
\includegraphics{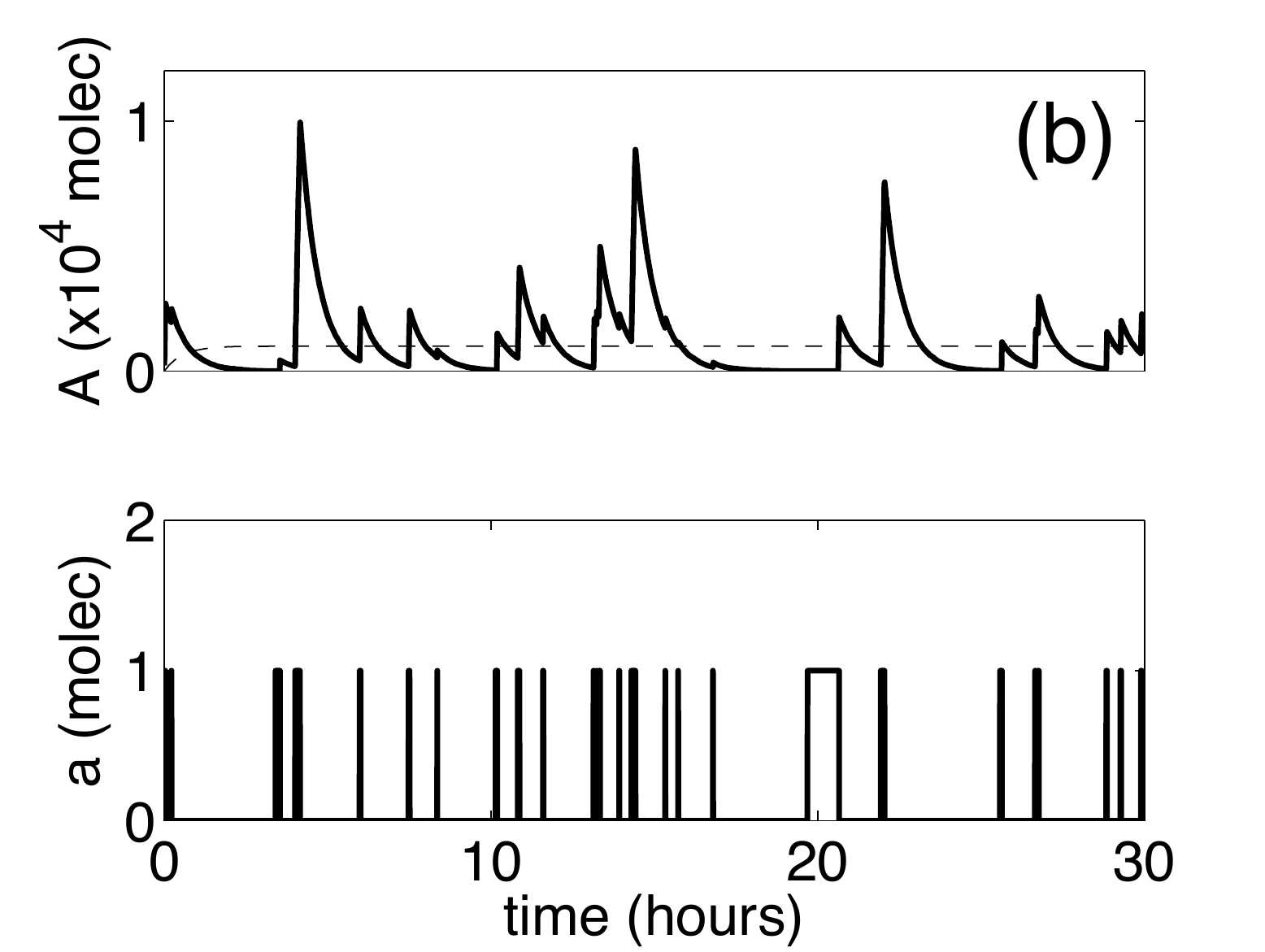}}}
\caption{Stochastic expression of a constitutive gene.
In the two panels, the upper row shows the temporal dynamics of
the protein level, and the lower row that of the mRNA level. In the top
row, the thin dashed line represents the solution of the deterministic 
(continuous) limit. Panels (a) and (b) correspond to the limits of
low (a) and high (b) noise level.
The parameter values are $k_{\rm deg,a} = 5\cdot 10^{-5}$~s$^{-1}$,
$k_{\rm deg,A} = 0.01$~s$^{-1}$, and
(a) $k_{\rm tx} =  0.025$~nM/s, $k_{\rm tl} =  0.2$~nM/s,
(b) $k_{\rm tx} =  2.5\times 10^{-4}$~nM/s, $k_{\rm tl} =  20$~nM/s.
}
\label{fig:stoch}
\end{figure}

Figure~\ref{fig:stoch} shows the stochastic dynamics of this system for two specific sets
of parameters, representing the limits of low (a) and very high (b) noise. In the case of
small noise levels, the number of protein molecules fluctuates around its deterministic steady
state (dashed line) with small-amplitude fluctuations. Such an expression
level corresponds to having around 2-3 molecules of mRNA continuously present in the
cell [a number that is subject to substantial fluctuations, as shown in the lower plot of
Fig.~\ref{fig:stoch}(a)], from which translation occurs in a more or less continuous manner.
The limit of large noise, on the other hand, arises when the transcription rate is extremely
small, so that usually there is only one mRNA or none present at any given time, as
shown in the lower plot of Fig.~\ref{fig:stoch}(b). In these
conditions, and if the translation rate is sufficiently high, protein production occurs in
bursts, leading to a highly irregular evolution of the protein level, as shown in the upper
plot of Fig.~\ref{fig:stoch}(b). Since this model is linear, it can be solved exactly,
as discussed in \cite{Friedman:2006p237} (see also \cite{Paulsson:2000p1098} for
a related calculation). The resulting distribution of protein levels is a Gamma
function with qualitatively different shapes for the two cases shown in Fig.~\ref{fig:stoch}
\cite{Friedman:2006p237}.
{\em Translational bursting} has been observed
experimentally in bacteria via both enzymatic \cite{Cai:2006p1051} and fluorescent-protein
\cite{Yu:2006p1056} markers (see Sec.~\ref{sec:exp} for a discussion of experimental
methods). Noticeably, the analytically obtained Gamma distributions in protein levels are
asymmetric, which qualitatively agrees with experimental observations
\cite{Ozbudak:2002p1237,Bar-Even:2006uq,Newman:2006p1239}. This provides
further support to the analytical considerations made above.

Interestingly, studies of stochastic gene expression in eukaryotic cells
\cite{Blake:2003p1242,Raser:2004p1244,Becskei:2005p1243,Golding:2005p1249,Blake:2006p1246}
yielded results which were not compatible with the translational bursting scenario
described above, but rather with a situation of {\em transcriptional bursting}, in which
it is the mRNA, and not the protein, that is being produced in
bursts. This behavior is compatible with a situation in which the DNA promoter undergoes
transitions between active and inactive states, during which the gene is being transcribed
or it is not, respectively. This suggests the presence in eukaryotes of transcriptional regulation
mechanisms in addition to those found in prokaryotes \cite{Raj:2008p561}, but the nature of
these mechanisms is still open to debate.

\subsubsection{Spatio-temporal stochastic description}

All the methods proposed so far have ignored the spatial distribution of biochemical
species inside the cell. Certainly, taking into account, in a model of gene regulation, the
tremendous structural intricacy of a cell's interior would be an extremely arduous
(and probably unnecessary) task. However, it is not so clear that one can ignore the
effect of the spatial distribution of biochemical species with respect to their binding targets,
when considering the time scales in which reactions take place inside the cell. With
this in mind, in recent years different groups have developed generalized stochastic
algorithms that allow the simulation of the dynamics of genetic circuits, taking into
account that biomolecules are reacting inside a spatially extended system.
The molecular motion is usually allowed to be governed by diffusion, and
the cytoplasm is normally considered a homogeneous medium. 

Two main approaches have been considered so far. In the first approach, space is discretized
and a reaction-diffusion master equation is written down and integrated
\cite{Hattne:2005p1253,Isaacson:2006p1255,Erban:2009p1252}. In the second description,
the motion of individual particles is described in terms of Brownian dynamics, and
reactions are accounted for when molecules collide within a certain range
\cite{Andrews:2004p1254,Erban:2009p1252}. A combination of both approaches,
leading to a more efficient algorithm and based on the so-called Green's function
reaction dynamics, was proposed in \cite{vanZon:2005p1251}.

\subsection{Experimental tools}\label{sec:exp}

\begin{figure}[htb]
\centerline{\resizebox*{12cm}{!}{\includegraphics[clip]{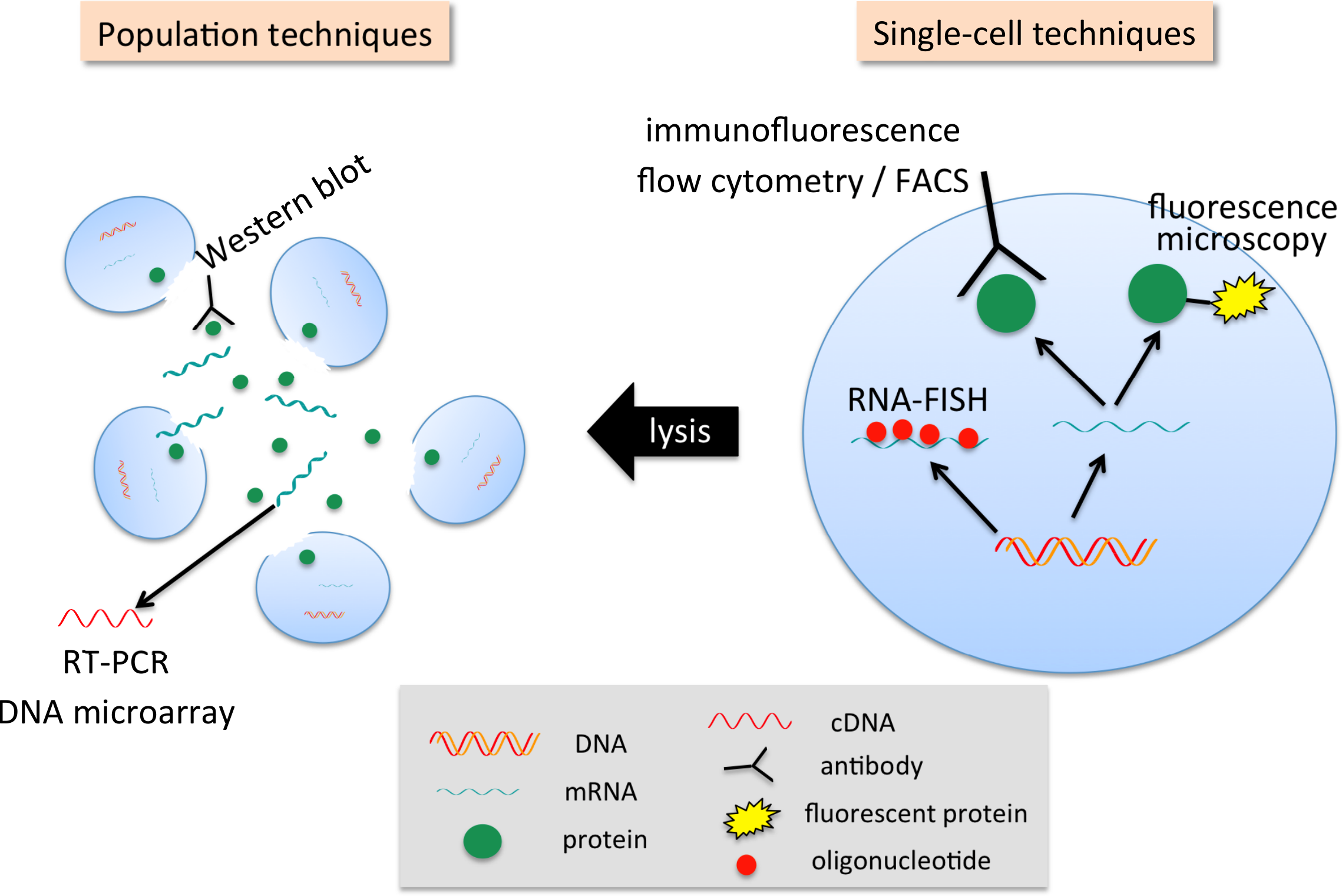}}}
\caption{Scheme representing the different experimental techniques for monitoring gene
circuit dynamics described in the text.
}
\label{fig:exp_cartoon}
\end{figure}

We now compare the different experimental methods currently available for monitoring
the dynamical behavior of genetic circuits. A scheme representing the different techniques is
shown in Fig.~\ref{fig:exp_cartoon}.

\subsubsection{Population techniques}

The molecular biology revolution of the past 40 years has provided us with an
excellent knowledge of the identity of the genes and proteins that underlie most cellular
processes, and in many cases of the interactions between them as well. Interestingly, a
large fraction of that information has been
obtained with population-level techniques, which are not able to discern individual cells.
{\em Western blotting}, for instance, is routinely used to detect the presence of a given protein
in tissue or cell extracts, where cells have to be lysed and their contents mixed with all
other cells in the sample, thus destroying the cells' individuality. Protein detection is made
in two steps: first, the protein of interest is separated according to its size via
{\em gel electrophoresis}; and second, an antibody specific to the protein and containing
an appropriate marker is applied to the sample to detect the amount of protein present
\cite{Renart01071979}.
The process has to
be repeated for a different cell population at different time instants, if one wants to
track the dynamical behavior of the system. This makes the process rather time-consuming
for monitoring the dynamics of genetic circuits.

Similar constraints are faced by
{\em reverse-transcription polymerase chain reaction} (RT-PCR), a widely used method
to detect RNA presence in a cell sample \cite{Singer-Sam11031990}.
The technique uses the enzyme reverse
transcriptase to generate the complementary DNA (cDNA) corresponding to the RNA molecule
of interest, and amplifies it via standard PCR. {\em Real time RT-PCR} uses then
fluorescent dyes that attach to the DNA, to provide an accurate (and very sensitive)
quantification of the original amount of RNA present in the sample \cite{Bustin:2000p1360}.
As in the case of
western blots, cells are lysed in the PCR process, so that the method is not able
to monitor the behavior of single cells. The temporal resolution of the measurements is
again naturally limited.

In the two methods described above, it is difficult to scale up the number of different species
that can be probed simultaneously. A technique that allows us to monitor the expression
of a very large number of genes simultaneously is based on the use of {\em DNA
microarrays}. These are chips containing thousands of DNA probes that hybridize
cDNA molecules corresponding to RNA molecules of interest \cite{Pollack:1999p1361}.
Hybridization of a given cDNA thus indicates that the matching RNA is present in the sample. Again,
however, cells have to be destructed to generate the sample, so that this technique
is not able to account for expression in single cells, nor it is
fit for dynamical measurements.

\subsubsection{Single-cell techniques}

None of the techniques discussed so far is able to quantify the amount of expression
exhibited by single cells. This problem can be solved via {\em immunofluorescence}.
This technique uses antibody staining,
discussed above in the context of Western blotting, in samples of intact cells,
which have to be fixed and permeabilized, so that the fluorescencently-tagged
antibody can reach its target inside the cell \cite{Lazarides:1974p1362}.
This kills the cells and prevents tracking
them in time, but allows us to quantify, via microscopy, the amount of protein
present in a relatively small number of cells (those that fall within the field of view
of the microscope). Several proteins of interest can be monitored simultaneously
(limited by the availability of antibodies and the cross-talk between the emission
bands of the fluorophores). A technique with similar capabilities that has been employed
profusely in recent years is {\em single-molecule RNA fluorescence in situ hibridization}
(RNA-FISH) \cite{Femino:1998p1363}. In this case, multiple short fluorophore-labeled oligonucleotides are
introduced in fixed cells. The oligonucleotides are designed so that they all bind
to different parts of the same RNA molecule. Cooperative binding of all the oligos
leads to a relatively strong fluorescence signal coming from a single RNA molecule.
In that way, single-molecule sensitivity can be achieved \cite{Raj:2008p1261}.
Again, cells must be
fixed and permeabilized, so that cell tracking in time cannot be performed.
On the other hand, the method is relatively simple to implement, since no
specific labeled antibodies need to be used, and genes for fluorescence proteins do not
need to be cloned into the cell's genome (see below).

In immunofluorescence and RNA-FISH, the number of cells that can be tracked
is limited by the size of the field of view of the microscope. A single-cell technique
that allows the quantification of protein expression of a very large number of
cells is {\em flow cytometry}. In this widely used technique, cells suspended
in a liquid flow pass through a laser beam. The light scattered from each cell
provides information about its morphology, while fluorescent light (emitted
by probes that the cell might contain) gives information about its chemical state
(such as the expression level of a labeled protein, for instance). Since a very
large number of cells pass through the laser beam every second, single-cell
information about a very large number of cells can be compiled in a short 
amount of time. The number of proteins that can be probed depends on
the number of tags that the cell contains, and is limited again by the
spectral cross-talk between the fluorophores. The efficiency of the technique
can be enhanced by allowing for the simultaneous analysis of multiple
samples via a ``bar-coding'' method. In this method, different samples are labeled
with distinct intensities of a fluorophore (the barcode) and mixed together,
prior to antibody staining and analysis by flow cytometry. After the analysis, the
samples can be separated from one another by the fluorescence intensities of their barcodes.
In that way, tens of samples can be analyzed in a single experiment,
which greatly reduces the amounts of antibody required
\cite{Krutzik:2006p287}. Also, adding an electrostatic deflection system to the
devices, cells can be sorted depending on their fluorescence, a technique that
is known as {\em fluorescence-activated cell sorting} (FACS)
\cite{Herzenberg:2002p1364}.

Due to the high number of cells that can be analyzed by flow cytometry
in a single run, that method is able to generate in a relatively straigthforward way
distribution functions of the state of a cell population, relative to the expression
of a given protein. In that way, it can be established whether a genetic circuit
has a bistable behavior or a simpler monostable one, for instance. 
However, as in all techniques discussed above,
this method is not intrinsically prepared to provide dynamical information,
since different cell populations must be used for different time points, and
the cells cannot be individually tracked.

All the techniques described above have been tremendously useful over the
years in providing information about the molecular determinants of cellular
function. However, as mentioned in the introduction and described in detail
below, many cellular processes are dynamical and subject to random
fluctuations. Thus usually cells in a population behave differently from
one another, and from time to time. An experimental technique that allows us
to track the biochemical state of single cells over time
is {\em time-lapse fluorescence microscopy}.
This technique heavily depends on {\em fluorescent proteins}, which
contain a chromophore that, when excited by light within a certain wavelength
range, emits fluorescent light of a higher wavelength. Many different types of
fluorescent proteins, covering a large part of the visible spectrum, have
been developed over the years \cite{Shaner:2005p723}. Tagging
the elements of genetic circuits with fluorescent proteins and monitoring a cell
population with a motorized microscope (with appropriate sets of filters to
separate excitation and fluorescent light), allows us to follow individual
cells in time and measure the state of the genetic circuits of interest
\cite{Locke:2009hc}. Single-molecule resolution can be achieved \cite{Yu:2006p1056}
by fusing the fluorescent protein to a membrane-targeting peptide,
which reduces the mobility of the fluorophore and thus prevents spreading
of the fluorescence signal through the cytoplasm, which happens when the
protein is mobile. Time-lapse fluorescence microscopy
allows for a relatively high temporal
resolution (on the order of seconds to minutes, depending on the
mechanical and optical characteristics of the microscope, and on the
potential photobleaching that the cells undergo), and permits the simultaneous
monitoring of multiple fluorescent markers (limited again by the cross-talk
among them). The number of cells that can be tracked is limited by the
field of view of the microscope. A comparative summary of the
features of the different experimental techniques discussed so far is provided
in Table~\ref{table:exp}.

\begin{table}[htb]
\caption{Characteristics of dynamical monitoring techniques for gene regulation processes
\label{table:exp}}
\begin{center}
{\begin{tabular}{|c|c|c|c|c|c|}\hline
   Technique  & Temporal & Single & Cell  & Number & Number   \\
    & resolution & cells & tracking  & of probes & of cells  \\\hline
   Western blotting & low & no & no & low & high   \\\hline
   Real-time RT-PCR & low & no & no & low & high   \\\hline
   DNA microarray & low & no & no & high & high  \\\hline
   Immunofluorescence & low & yes & no & medium & low  \\\hline
   Single-molecule RNA-FISH & low & yes & no & medium & low  \\\hline
   Flow cytometry & low & yes & no & medium & high  \\\hline
   Fluorescence microscopy & high & yes & yes & medium & low  \\\hline
  \end{tabular}}
\end{center}
\end{table}

There are different ways to label the activity of a genetic circuit with a fluorescence
protein. The most logically direct way is by fusing the fluorescent protein with
the protein to be monitored. Such {\em protein fusion}
provides a fluorescence signal
that is a direct measure of the protein level. However, if the protein being monitored
is supposed to have a functional effect in the circuit (as is usually the case, for instance
regulating the transcription of a gene or activating another protein enzymatically),
it is necessary to check that the fluorescence tag that has been attached does
not disrupt the function of the protein, something that is not always easy to establish.
A different, much less ``invasive'' labeling method is to add 
to the cell (in the case of a bacterium, either into its chromosome or on a plasmid),
an extra copy of the promoter fused only to the fluorescent protein. In that way,
the fluorescence signal becomes a measure of the activity of the promoter that
controls the expression of the protein of interest: when that protein is being expressed,
the fluorescence signal will increase. Mathematically, we can describe the dynamics
of the fluorescent protein by means of the following expression:
\begin{equation}
\frac{dG}{dt} = f(\{P\})-\delta G\,,
\label{eq:gfp}
\end{equation}
where $G$ stands for the concentration of green fluorescent protein (GFP), and
$\{P\}$ represents the set of all transcription factors affecting the expression
of the promoter from which GFP is expressed. This promoter is identical to the one in the genetic
circuit whose activity we want to measure. The negative term
corresponds to GFP degradation. Assuming no differences in the translation of GFP and
the protein of interest from their respective mRNA, we can identify the real promoter activity
of the protein of interest as $P_A=f(\{P\})$, and from Eq.~(\ref{eq:gfp}) we get to the following
rule to obtain the promoter activity from the fluorescent signal $G$ being measured:
\begin{equation}
P_A = \frac{dG}{dt} +\delta G\,.
\label{eq:promact}
\end{equation}
The first term in the right-hand side of this equation can be obtained by simply deriving the
fluorescence time series. Thus, provided one can measure the degradation rate of the
fluorescent protein in a separate experiment (by monitoring the decay of the fluorescence
signal following a halt in the production of the protein), we can easily calculate the promoter
activity of the protein of interest.

Time-lapse fluorescence microscopy has two main limitations. First, when monitoring
rapidly dividing cells, such as bacteria or stem cells,\footnote{Here, ``rapid'' means that
the division time is smaller than the tipical timescale of the genetic circuit of interest.}
the field of view fills up with cells rather quickly, at which point the natural behavior of
the cells, and/or our ability to monitor them, are disrupted: mammalian
cells, for instance, might start to contact and signal one another, possibly changing
their mutual behavior; bacteria, on the other hand, might start growing on top of each other
and we would lose our ability to discern them. Therefore, the standard time-lapse
fluorescence microscopy setup has a strong time constraint on the maximum duration
of the experiment. A second problem is that changing the conditions to which the cells
are subject in real time is not a simple task. These two problems have been solved
by growing the cells in {\em microfluidic} chambers, which can get rid of overflowing cells
and allow the addition or removal of nutrients and other signals in real time
\cite{Bennett:2009kh}.
This setup also enables us to maintain constant the conditions to which the cells are
subject, something that frequently does not hold in standard microscopy, since
while cells grow the medium conditions change. Combining
microfluidics with time-lapse fluorescence microscopy thus provides a way
of monitoring the dynamical behavior of genetic circuits for long times in well-controlled
conditions. In order to obtain a better control over the cells, {\em optical tweezers} can be
used. These devices use the radiation pressure exerted on cells by highly focused
laser beams to trap the cells \cite{Ashkin:1987p1359,volpe:231106}, deform them \cite{Rao:2009fk},
position them \cite{Timp:2009p235,Eriksson:2010p500},
and even assemble them into tissue-like structures \cite{Mirsaidov:2008p1358}. The combination
of fluorescence microscopy with microfluidics and optical trapping thus provides an ideal
experimental setup, with which to monitor the dynamical behavior of genetic circuits with
high temporal resolution and for large observation times.

\section{Dynamics of gene regulation}\label{sec:dyn}

In this section, we present an overview of the experimental evidence of
dynamical behavior of genetic circuits, concentrating on examples of how this feature can
be used to identify the molecular mechanisms underlying the corresponding cellular
functions.

\subsection{Oscillations}\label{sec:osc}

In the last decades, an overwhelming amount of evidence has shown
that the biochemical state of cells is far from being stationary, and that this dynamics has
frequently a functional role. The simplest case of dynamical behavior is oscillatory
activity\footnote{The term ``oscillation'' is being used here to refer to {\em periodic} oscillations.}, which,
as shown in Sec.~\ref{sec:2d} above, is associated with limit-cycle behavior in phase space.
A large number of examples of oscillatory activity in biological systems have been
identified in the last 30 years, with periods ranging from tens of milliseconds (as in the case
of electrical pulsing activity in neurons) to months (in seasonal rhythms) \cite{goldbeter}.
Given that nonlinear physics tells us that no more than two degrees of freedom (or even one
in the presence of delays) are required to generate oscillations \cite{Strogatz:1994fk},
it is reasonable to expect that periodic cellular activity can be generated by relatively small-sized
genetic circuits. One instance of such a system, namely an activator-repressor circuit, was
given in Fig.~\ref{fig:2d}. In the following paragraphs, different examples of oscillatory
cellular processes will be given, most of which can be associated with genetic circuits with
small numbers of components.

One of the first examples of dynamical behavior at the molecular level that was identified
and characterized in detail was {\em circadian rhythmicity}. Living organisms exist in an
environment naturally subject to periodic cues, for instance from the light-dark cycle of earth's rotation around its
axis. As a result, it was evolutionary beneficial for living systems to develop biochemically-based
mechanisms that allowed them to anticipate the environmental rhythms. Circadian clocks have been identified in a wide
variety of organisms, with very different molecular circuitry in most cases\footnote{In simple
organisms such as cyanobacteria, circadian rhythms can in fact arise in pure protein
circuits, without the need of transcriptional regulation \cite{Nakajima:2005p1278}.}
\cite{Young:2001p1271}. The first molecular models of genetic circuits underlying circadian
rhythmicity focused on small circuits, containing the main genes that were known to
be relevant for the phenomenon \cite{Goldbeter:1995p1272,Scheper:1999p1273}.
These early models already identified circadian oscillations as instances of limit-cycle
behavior. As new facts and new genes were being identified the models became more
complex \cite{Leloup:2003p1274}, although predictions based on design principles
derived from our knowledge of simple genetic circuits are still being made
\cite{Kuczenski:2007bh}.

A second situation in which periodic behavior is apparent is the cell cycle.
As in the case of the circadian clock, many phenomenological models relaying on limit
cycle attractors were proposed early on, well before most of the molecular components
underlying the cell cycle were identified \cite{goldbeter}. Recent years have witnessed a
substantial increase of experimental knowledge about the cell cycle of
a range of organisms including bacteria 
\cite{Biondi:2006p839,Collier:2007p211}, yeast \cite{Orlando:2008p296,Kaizu:2010p1124}, 
amphibian embryos \cite{Pomerening:2005p1275}, and mammalian \cite{SakaueSawano:2008p539}
(in particular human \cite{Pomerening:2008p1276}) cells,
among many others. Using that information,
molecular models with various levels of detail have been built
\cite{Tyson:1999p977,Tyson:2001p1277,Sveiczer:2000p853}.

\begin{figure}[htb]
\begin{center}
\resizebox*{5cm}{!}{\includegraphics{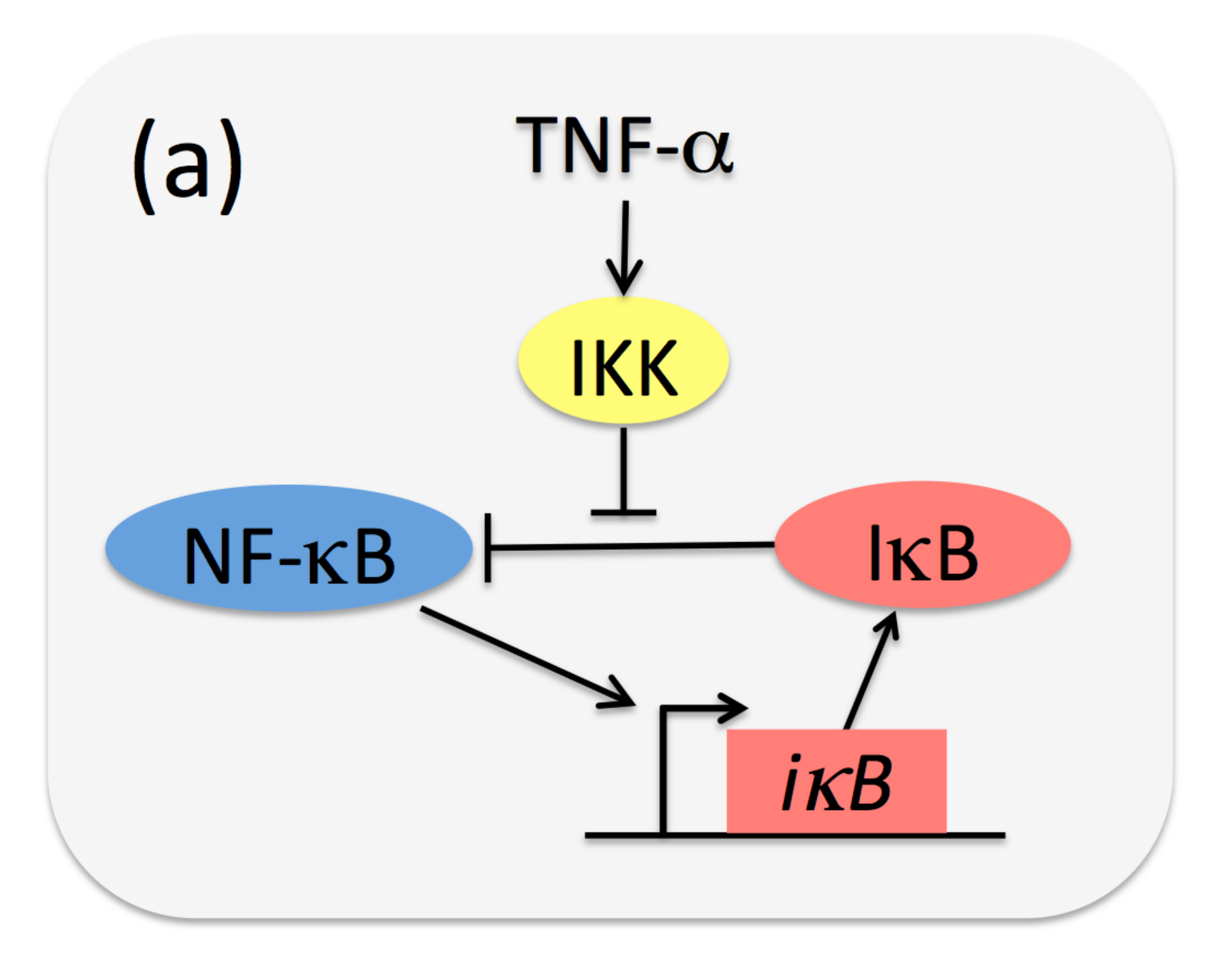}}~~~
\resizebox*{5.3cm}{!}{\includegraphics{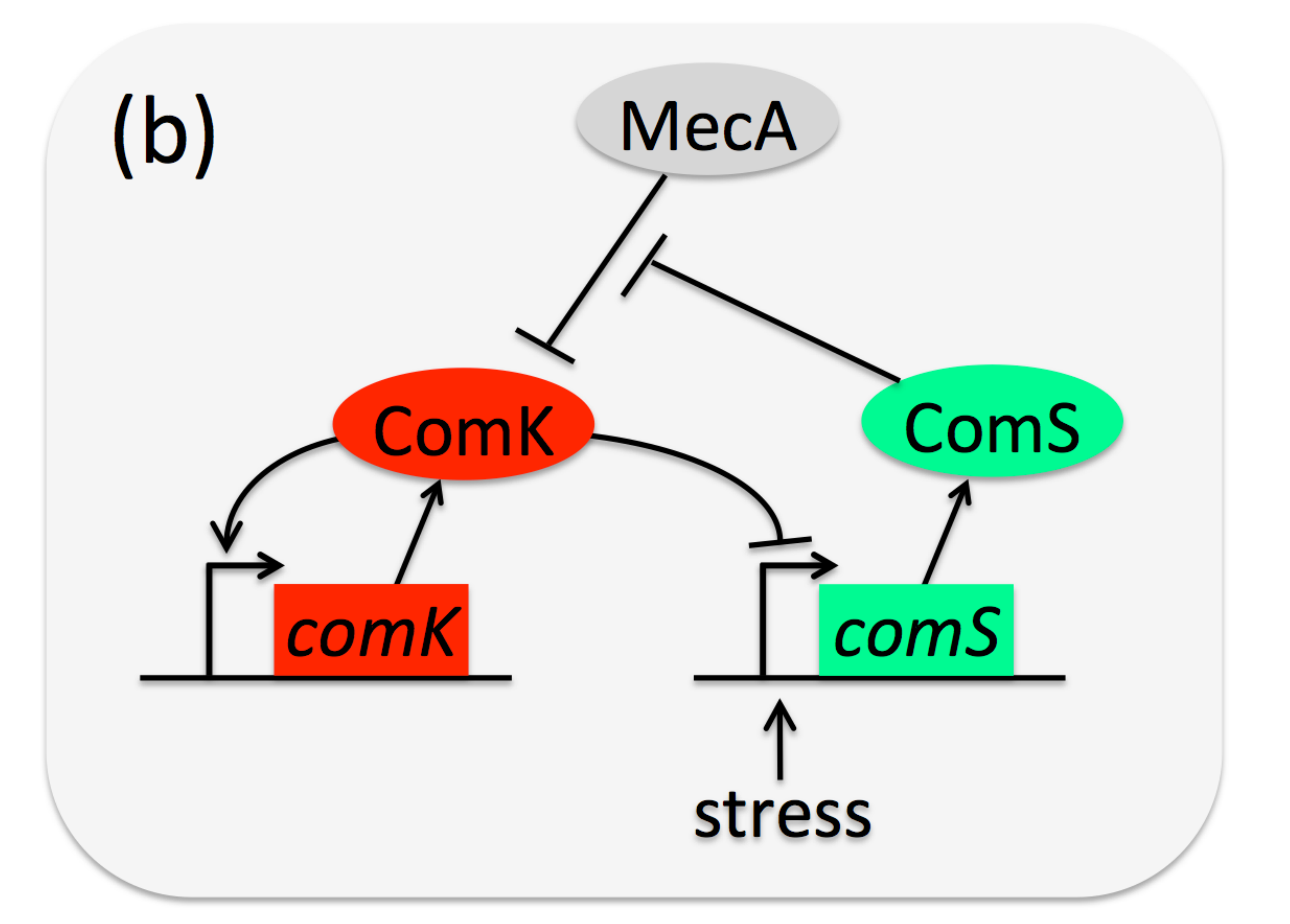}}
\end{center}
\caption{Natural genetic circuits leading to oscillatory (a) and excitable (b)
dynamics. The scheme in (a) is a simplified view of the circuit underlying
periodic oscillations in the activity of the transcription factor NF-$\kappa$B. That
protein activates transcription of I$\kappa$B, which binds and deactivates NF-$\kappa$B
itself. The resulting negative feedback loop is subject to delays naturally associated with the
cellular processes taking part in it (transcription, entry into and exit from the nucleus, protein
binding), and leads to oscillatory behavior. The circuit is affected by the cytokine TNF-$\alpha$
through its activation of the kinase IKK, which interferes with the inhibition of NF-$\kappa$B
by I$\kappa$B by phosphorylating the latter. The scheme in (b) represents the core genetic
circuit underlying competence in {\em B. subtilis} bacteria under stress. Here the master regulator
of competence, ComK, is activated by the stress-sensing protein ComS, which interferes with
the degradation of ComK by a protease complex controlled by MecA. This triggers the
positive feedback loop to which ComK is subject, and competence is initiated. Inhibition
of ComS production by ComK later terminates the competence excursion, leading to
excitable dynamics.
}
\label{fig:dyncirc}
\end{figure}

Signaling networks have also been shown to exhibit oscillatory activity. In 2002, population
studies suggested that the transcription factor NF-$\kappa$B, which regulates multiple
cellular processes including cell proliferation, apoptosis and inflammation, responds in
a periodic manner to stimulation by the cytokine TNF-$\alpha$ \cite{Hoffmann:2002p349}.
In that same work, those experimental observations were used to propose a molecular model
of a circuit [Fig.~\ref{fig:dyncirc}(a)] involving NF-$\kappa$B, its inhibitor I$\kappa$B, and the
kinase IKK, which phosphorylates
I$\kappa$B, targeting it for degradation and thus activating NF-$\kappa$B.
Subsequent single-cell analyses via
fluorescence microscopy confirmed the oscillatory behavior at the level of individual
cells \cite{Nelson:2004zt}. The observed period of the oscillations was around 2~hours,
for the cytokine levels tested. When this circuit was stimulated in a pulsed manner, instead of
receiving a constant level of cytokine, the pattern of oscillations in NF-$\kappa$B activity
was seen to depend in a non-trivial way on the stimulation frequency \cite{Ashall:2009il}.
This suggests a potential
role of the oscillations as information processors, taking into account that cytokine signaling
in inflammatory tissues might be time-dependent.

A different study made use of a microfluidics
setup to vary in a controlled manner the level of (constant) cytokine signal. A single-cell
analysis elegantly revealed that as the amount of signal decreases, it is the number of
responding cells, but not the average NF-$\kappa$B signal per pulse, that decreases
\cite{Tay:2010p1036}.
Therefore one can infer that the limit cycle dynamics disappears in a discontinuous manner
as the control parameter varies, which is consistent with a global bifurcation like the SNIC
shown in Fig.~\ref{fig:2d}(f), instead of a local bifurcation like the supercritical Hopf
displayed in Fig.~\ref{fig:2d}(e). It would be interesting to quantify how the frequency
of the NF-$\kappa$B oscillations varies as the control parameter changes across the bifurcation.

Other examples of oscillatory activity in signaling circuits include the signaling cascade
acting upon the key regulator of cell proliferation ERK \cite{Shankaran:2009p374},
and various pathways involving the STAT family of transcription factors, which play
an important role in the immunological response of cells, and are known to
interact with NF-$\kappa$B \cite{Yu:2009p332}. In this latter case, the evidence of
oscillatory activity, affecting the phosphorylated state of the STAT protein,
comes only from population studies \cite{Yoshiura:2007gb}.
STAT oscillations have a period of around 2~hours, as the ones in NF-$\kappa$B,
while oscillations in ERK nuclear translocation have a shorter period, of around
15~min. The functions of these two types of oscillations are unknown.

\subsection{Excitability and bursting}\label{sec:exc}

Periodic oscillations are not the only type of dynamical behavior exhibited by
genetic systems. The tumour suppressor protein p53, for instance, responds
to DNA damage by exhibiting pulses of activation, the number of which
varies from cell to cell within a population \cite{Lahav:2004p488}. It is still unclear
whether these pulses are part of an oscillatory response \cite{GevaZatorsky:2006p1269}, 
or an example of excitable dynamics such as the one shown in Figs.~\ref{fig:2d}(b,d),
but the current evidence points to this second possibility \cite{Batchelor:2009p419}.
More likely, this system exhibits the two types of dynamical behavior, depending
on the condition to which the circuit is subject, as happens in the bifurcation
diagrams shown in Figs.~\ref{fig:2d}(e,f).
In any case, the dynamics exhibited by this system can be used to suggest
models of genetic circuits that might underlie the response of p53 to DNA damage
\cite{GevaZatorsky:2006p1269}.

A situation in which clear non-oscillatory dynamical activity has been reported
is the response to nutritional stress of the bacterium {\em Bacillus subtilis}.
The stress response of this organism is very rich, including terminal differentiation
into a spore, which is a dormant and environmentally resistant state, and reversible
differentiation into a state of {\em genetic competence}, in which the cell cannot
divide and becomes capable of taking up exogenous DNA \cite{Grossman:1995p1279}. Morphologically,
the competence state is characterized by noticeable changes in the cell's buoyancy, 
which enables the separation of competent and vegetative cells in population-level
studies. In that way, much was learned about the biochemistry and the molecular
biology of this process (which is very interesting for its genetic engineering
applications, since it provides a way of introducing DNA into cells), without having
to see what happens to individual cells.

Time-lapse fluorescence microscopy,
on the other hand, promptly revealed that the reversible differentiation into
competence was strongly dynamical at the single-cell level, and characterized
by transient pulses of activity of the master regulator of competence ComK,
a transcription factor that activates the expression of more than 100 genes that
determine the transition to the state of competence. Initiation of competence
is stochastic, with only a small fraction (around 5\%) of all stressed
cells accessing to it (while the rest sporulate or die, due to nutrient limitation).
However, once the cells enter into the competent state, they exit from it in a deterministic
manner: the time spent in competence is much less variable than the initiation time of
different cells. This is the classical trademark of excitability,
as represented in Fig.~\ref{fig:2d}(b). A systematic investigation of the genetic
network known to be involved in the development of competence \cite{Hamoen:2003p1280}
revealed that a small genetic circuit was responsible for the dynamical process
underlying the transient differentiation into competence \cite{Suel:2006oq}.
The circuit, shown in Fig.~\ref{fig:dyncirc}(b), contains the master regulator of competence,
ComK, the stress-monitoring
protein ComS, and the protease MecA, which degrades both ComK and ComS in
a competitive manner.

\begin{figure}[htb]
\centerline{\resizebox*{14cm}{!}{\includegraphics[clip]{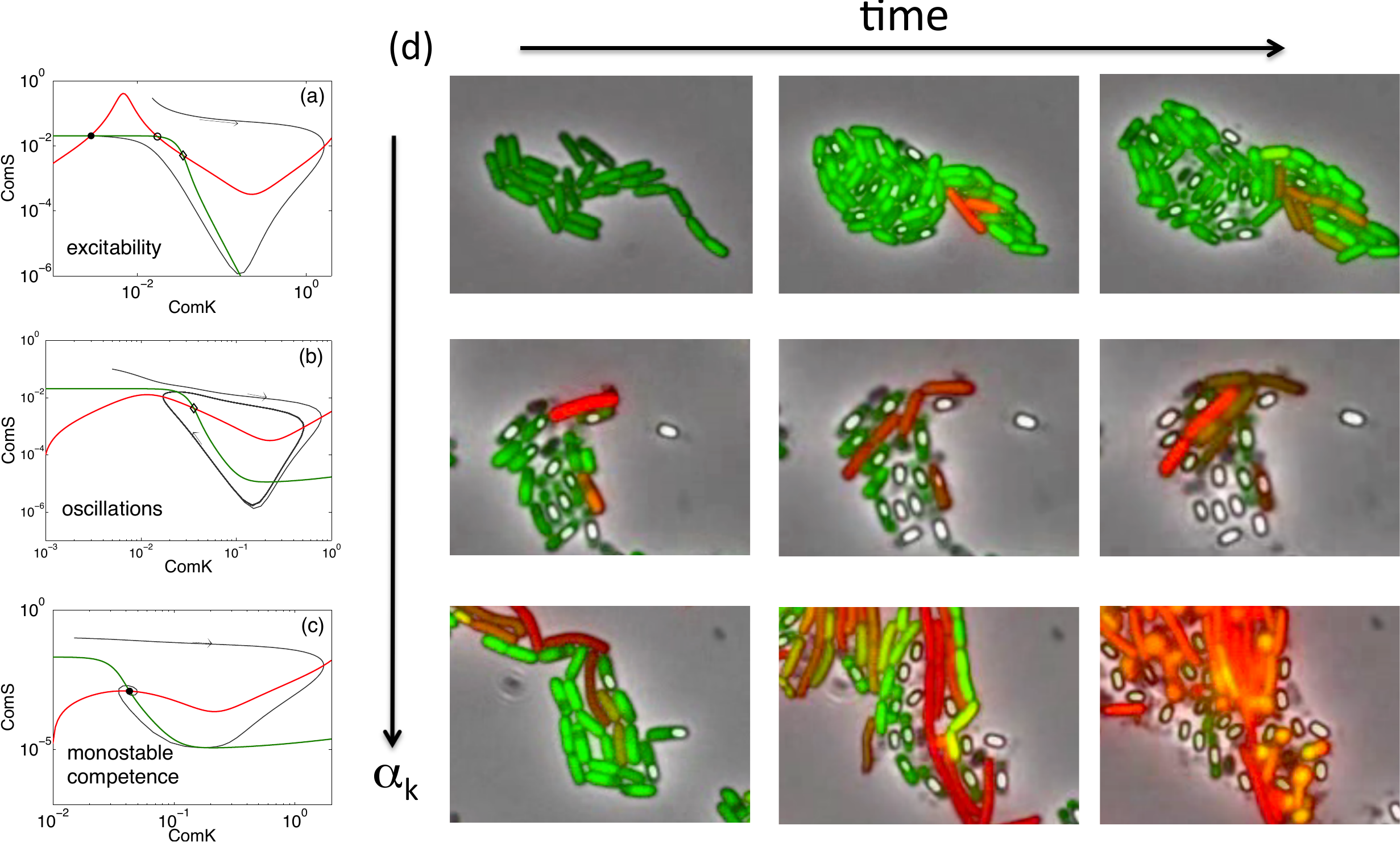}}}
\caption{Bifurcation analysis of a genetic circuit {\em in silico} and {\em in vivo}. Panels
(a-c) shows phase portraits of the genetic circuit responsible for the transient
differentiation into competence [Fig.~\protect\ref{fig:dyncirc}(b)],
for increasing values of parameter $\alpha_k$, which corresponds to the basal
transcription of ComK. Different dynamical regimes are reached, from excitability (panel a),
which is the natural response to stress, to oscillations (panel b) and monostability at high
ComK levels (panel c). The corresponding experimentally observed behavior is
shown in the snapshots of panel (d), with time proceeding from left to right. Red (green) represents
the fluorescence protein being expressed from the ComK (ComS) promoter.
Experimental data kindly provided by G\"urol S\"uel.
}
\label{fig:bacillus}
\end{figure}

Applying the methods described in Sec.~\ref{sec:2d} above, the genetic circuit can
be described by two coupled ordinary differential equations for the proteins ComK and
ComS. This model exhibits, for moderate levels of stress (which enters the equations
as the parameter controlling the maximum expression level of ComS), excitable dynamics
that explains
the transient differentiation events into competence observed experimentally.
The corresponding phase portrait is shown in the upper left plot of Fig.~\ref{fig:bacillus}.
The snapshots on the top row of that figure show three time instants of the evolution
of a collection of {\em B. subtilis} cells subject to nutritional stress. The red and green
colors represent fluorescence levels obtained from promoter fusions. The white circles
in these snapshots highlight one cell undergoing competence.

Different predictions of this model were tested in order to validate the circuit. Among
them a bifurcation analysis was made using as a control parameter the basal expression
of ComK, denoted $\alpha_k$. This parameter can be tuned experimentally by adding,
via chromosome integration,
a second copy of the ComK gene under the control of an IPTG-inducible promoter.
Adding different amounts of IPTG to the medium in which the bacteria grow, one
can attain a continuous control of basal ComK expression. In parallel, a bifurcation
analysis of the model for increasing $\alpha_k$ can be performed, leading to the
conclusion that as the unregulated expression of ComK increases, cells tend
to a state of permanent competence (something to be expected, since eventually
the cells contain a high and constant amount of exogenous ComK), but interestingly
they do so by passing through a region of oscillatory dynamics, something that is
not present in this system under natural conditions. Experimental observations confirm
this prediction \cite{Suel:2007cr}, as shown in the snapshots of the middle row of
Fig.~\ref{fig:bacillus}. The white circles in this case highlight a cell that (together
with its daughters, since the duration of the pulse, around 20 hours, is
longer than the cell cycle time) undergoes two consecutive entries into competence.
The state of monostable competence (bottom row of Fig.~\ref{fig:bacillus})
in which all cells end up being competent or spores, is also un-natural in these cells.

Other perturbations can be applied to the circuit, and can be used to continue validating the
model. In fact, both experiments and theory agree that in this system the duration and
frequency of the competence pulses can be controlled independently of one another,
by tuning the basal expression rates of ComK and ComS, respectively \cite{Suel:2007cr}.

Excitable dynamics is being identified in other cellular systems. In particular, recent evidence
points to its involvement in the maintenance of pluripotency in embryonic stem cells.
In fact, recent studies have revealed two intriguing features of pluripotency. First, there
is substantial heterogeneity in the expression levels of certain proteins known to be
relevant for the development of pluripotency, such as Nanog, while the variability is
much smaller in other key players of the process \cite{Chambers:2007p622}. Second,
Nanog expression seems
to be bimodal in stem cell populations, with most of the (pluripotent) cells expressing
a high level of Nanog, while a relatively small fraction of the cells express Nanog
at low levels. These latter cells have been shown to be closer to differentiation
\cite{Kalmar:2009vn}. More intriguingly, low Nanog cells have been seen to revert
back to a high Nanog state after some time \cite{Singh:2007p346}. These characteristics
can be explained by excitable dynamics, which naturally leads to (i) a certain level
of variability, (ii) a non-trivial distribution of Nanog expression, with a certain degree of
bimodality (with the more populated state corresponding to the stable state of the system,
and the less populated one related with those cells that are in the excited branch of
the phase plane), and (iii) spontaneous dynamical transitions between the two states,
driven by noise. Under these starting hypothesis, we set out to examine the possibility
that the genetic network underlying pluripotency might contain a genetic circuit that
provided a mechanism for excitability. A small genetic circuit containing
Nanog and another protein relevant for pluripotency, namely Oct4, was seen to
allow for an excitable topology in phase space. It is appealing to conjecture that
the pluripotent state in embryonic stem cells is excitable in order to have the cells
primed for differentation in the embryo, when needed along the development process.
This would provide an important functional role for excitability in cell dynamics.

Recently, another strongly dynamical (but non-periodic) behavior has been uncovered
that could constitute a new means of transcriptional regulation. The evidence 
comes from the calcium stress response of yeast cells \cite{Cai:2008bs}. In the
presence of calcium, the transcription factor Crz1 moves to the cell nucleus,
where it activates the expression of more than 100 genes involved in calcium
adaptation. Interestingly, time-lapse fluorescence microscopy revealed that
this nuclear translocation does not take place in a continuous manner, but
it does so in a pulsed manner, with rapid and short bursts of localization of
Crz1 into the nucleus \cite{Cai:2008bs}. Even though the molecular mechanisms
underlying these oscillations are not clear, due to the striking observation that
the stimulation controls not the amplitude of the localization bursts but
their frequency, it is attractive to think that cells use this frequency-modulation
method of encoding cellular signals as a way of reaching coordinated regulation
by promiscuous transcription factors, even when their different targets
have distinct input functions \cite{Cai:2008bs}. It can be expected that many
more pulsing and bursting phenomena will be found in cells in the upcoming
years, and will be seen to play relevant roles in cellular processes.
As an example, a recent work has revealed the existence of a refractory time
in transcription cycles in mammalian cells \cite{Harper:2011p1370}, below which
a second cycle cannot be elicited. Given that refractoriness
is one of the key features of excitable dynamics, it is appealing to conjecture that the dynamical
regime underlying the occurrence of stochastic transcription cycles might be
excitability.

\subsection{Natural vs synthetic circuits}

A decade ago, synthetic circuits showed us that small genetic circuits can
have well-defined and non-trivial effects in cells \cite{Sprinzak:2005p1224,Purcell:2010p1207}.
Examples include genetic
oscillators \cite{Elowitz:2000lh,Atkinson:2003p1217,Fung:2005p1287,Stricker:2008bd},
switches \cite{Gardner:2000p1285,Becskei:2001p1286,Atkinson:2003p1217},
stabilizers \cite{Becskei:2000p1284}, and counters \cite{Friedland:2009p716},
among many others. Engineering delays in a negative feedback loop have also
been recently shown to lead to oscillations with controllable period in animal
cells \cite{Swinburne:2008p1310}. Synthetic circuits are ideal
to test the hypothesis of the existence of core genetic circuits, because they
are relatively isolated from the rest of the cell, ideally sharing only the resources
(but not the regulation) with the rest of the cellular machinery.

More recently, however,
synthetic circuits have also began to help us understand the behavior and structure
of natural genetic circuits, by revealing the possible behaviors that a given
circuit architecture has. A recent example of this has emerged in the case of
bacterial competence discussed in the previous section. A comparison between
the phase planes shown in the top left plot of Fig.~\ref{fig:bacillus} and in
Fig.~\ref{fig:2d}(b), and their corresponding circuit architectures, indicates that
a similar dynamical behavior can be produced by different circuit topologies.
The question is then how does evolution lifts this functional degeneracy among
genetic circuits. Construction of an activator-repressor excitable system in
{\em B. subtilis} by reusing the same components of the natural competence
circuit (which was subsequently removed, leaving only the synthetic excitable
circuit in the cell) led to a functional synthetic competence module. Since the
two circuits (the natural and the synthetic) share most of the components
and their interactions, comparing their behaviors should reveal differences that
are due only to the circuit architecture. Interestingly, in this case the comparison
unveiled that the natural circuit was noisier than the synthetic one, in the
sense that the duration of the competence excursion was far more variable from
event to event in the former system than in the latter \cite{Cagatay:2009kx}.
A transformation assay showed that noise would be beneficial in the presence
of an uncertain environment, providing a reason for the occurrence in nature
of one of the two circuits over the other. Another systematic study along those lines
has been carried out in feedforward circuits in {\em E. coli} \cite{Kittisopikul:2010fu}.

A similar approach has been recently
employed to better understand the circuit responsible for the pulsing response
of p53 to DNA damage described in Sec.~\ref{sec:exc} above. The study
showed that certain features of the dynamical behavior of the system, namely
the amplitude, frequency and damping rate of the oscillations, can be controlled
by the stimulus level and the circuit topology. 

\section{Noise in gene regulation}\label{sec:noise2}

The experimental evidence presented in the preceding section usually displays
a substantial amount of noise and variability. Random fluctuations arise,
as argued in Sec.~\ref{sec:noise}, from the discreteness intrinsic to biochemical
reactions involving a small number of molecules, as is commonly the case in cells.
We now review several studies in which stochasticity in genetic circuits has
been taken into account, describing how noise is measured, how it can be
controlled to determine its effects, and what are those effects.

\subsection{Observing and measuring noise}

In the early works that quantified noise in bacteria \cite{Elowitz:2002pi,Swain:2002mi} a distinction
was made between {\em intrinsic} and {\em extrinsic} noise in gene regulation. Intrinsic
noise refers to the stochasticity inherent to the biochemical reactions underlying the
different processes involved in gene expression and regulation: for instance, the random
binding of transcription factors and RNA polymerases to DNA promoters, and of ribosomes
to mRNA molecules. Such fluctuations cause the expression of two identical promoters
placed in different
regions of the chromosome to express different levels of protein (even when there are no
differences between the transcription efficiencies associated with the location the promoter
within the chromosome -- which is accomplished in bacteria by placing the promoter
symmetrically with respect to the origin of replication of the chromosome). Extrinsic
noise, on the other hand, corresponds to fluctuations {\em in the number} of the external
species (i.e. those not subject to regulation by the elements of the genetic circuit under study)
involved in the gene regulation process (including, for instance, signaling molecules and
transcription factors).
In that way, extrinsic noise has no differential effect on the expression levels of the two identical promoters
described above: both will see the same fluctuations in the external components acting upon
them.

The distinction between intrinsic and extrinsic noise is reminiscent of one amply
considered in the field of statistical mechanics, between \emph{internal} and \emph{external} noise.
Even though the two pairs of concepts share some transversal features (e.g. extrinsic
noise is frequently parametric, acting upon the system parameters, as is usually
the case in external noise), there are certain differences. For instance, extrinsic noise,
due to its own nature, is strongly modulated by the cell size, and thus consistently
exhibits a temporal correlation in dividing cells \cite{Rosenfeld:2005fu}.
Thus extrinsic noise is usually colored, with a correlation time equal to the cell cycle
time. Frequently, this time is noticeably larger than other time scales of the cell,
and hence cannot be neglected. This contrasts with the external noise considered in statistical
physics, which can be assumed to be white under certain conditions. In any case, the
distinction between intrinsic and extrinsic noise makes biological sense, and has
helped us understand the role of fluctuations on genetic circuits. By way of example,
extrinsic noise has been identified as the source of qualitatively novel dynamics
\cite{Hasty:2000mb,Samoilov:2005p1250} due to its frequently multiplicative character,
which leads to systematic effects on the average dynamics of the system
\cite{nises} (although recently, qualitatively novel dynamics emerging from intrinsic
noise has also been identified \cite{Turcotte:2008ve,Rue2011}). Also, a recent theoretical study
has pointed out that the correlation between the two types of noise determines in
a relevant manner the information-carrying capacity of biochemical networks
\cite{TanaseNicola:2006p553}. It is also worth noting that
the standard Gillespie method for the discrete simulation of genetic circuits
incorporates in a natural way the intrinsic fluctuations, but not the extrinsic ones.
For a recent extension of the method that deals with extrinsic noise see for instance
\cite{Shahrezaei:2008p822}.

\subsection{Controlling noise}

\begin{figure}
\centerline{\resizebox*{14cm}{!}{\includegraphics[clip]{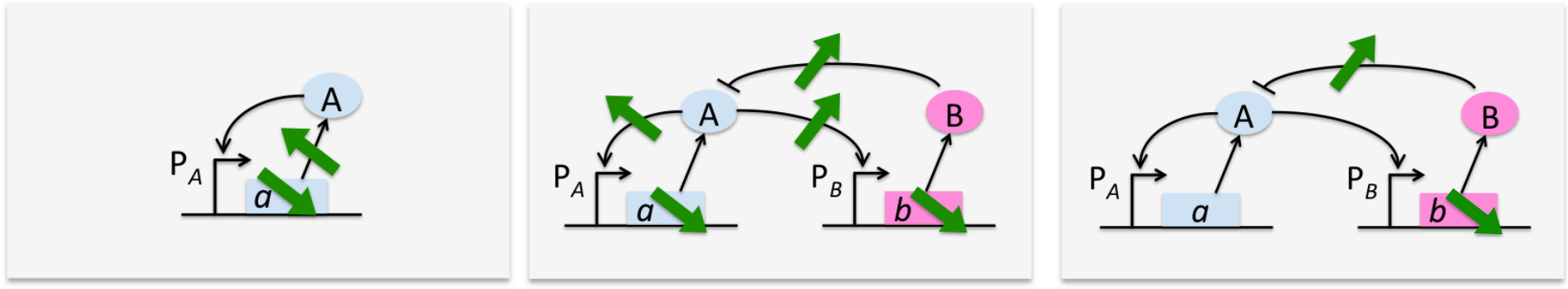}}}
\vskip3mm
\centerline{\resizebox*{14cm}{!}{\includegraphics{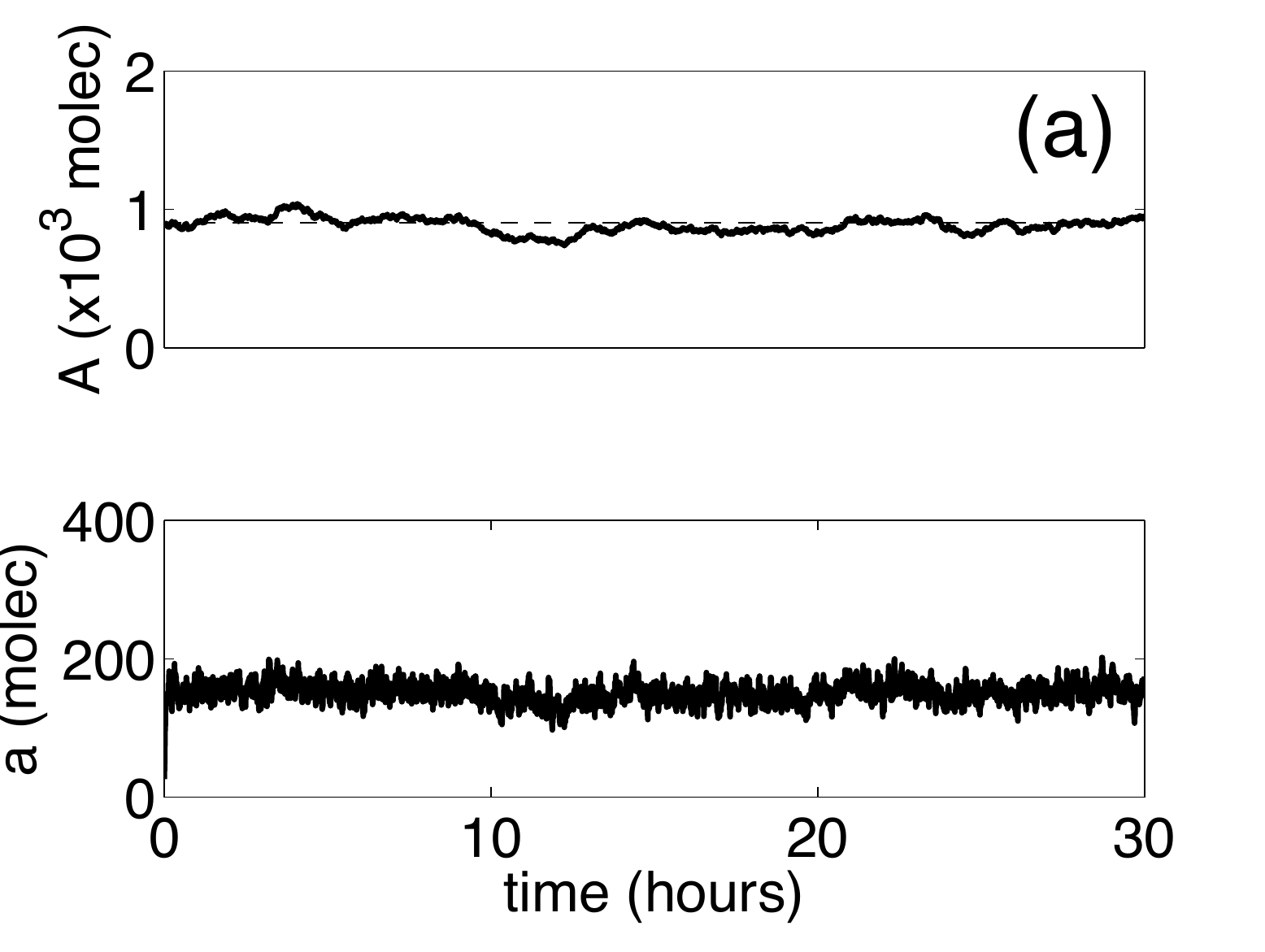}
\includegraphics{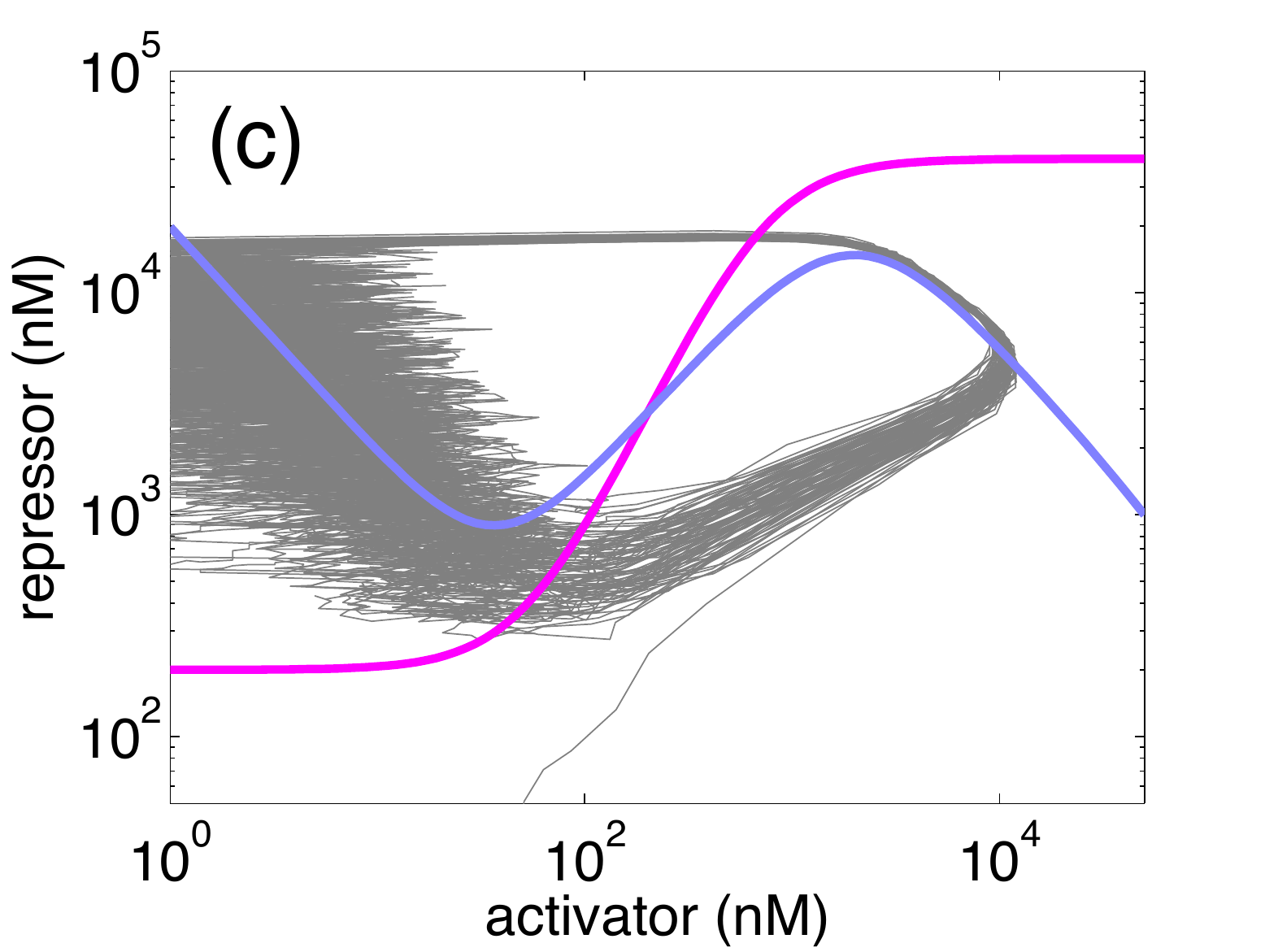}
\includegraphics{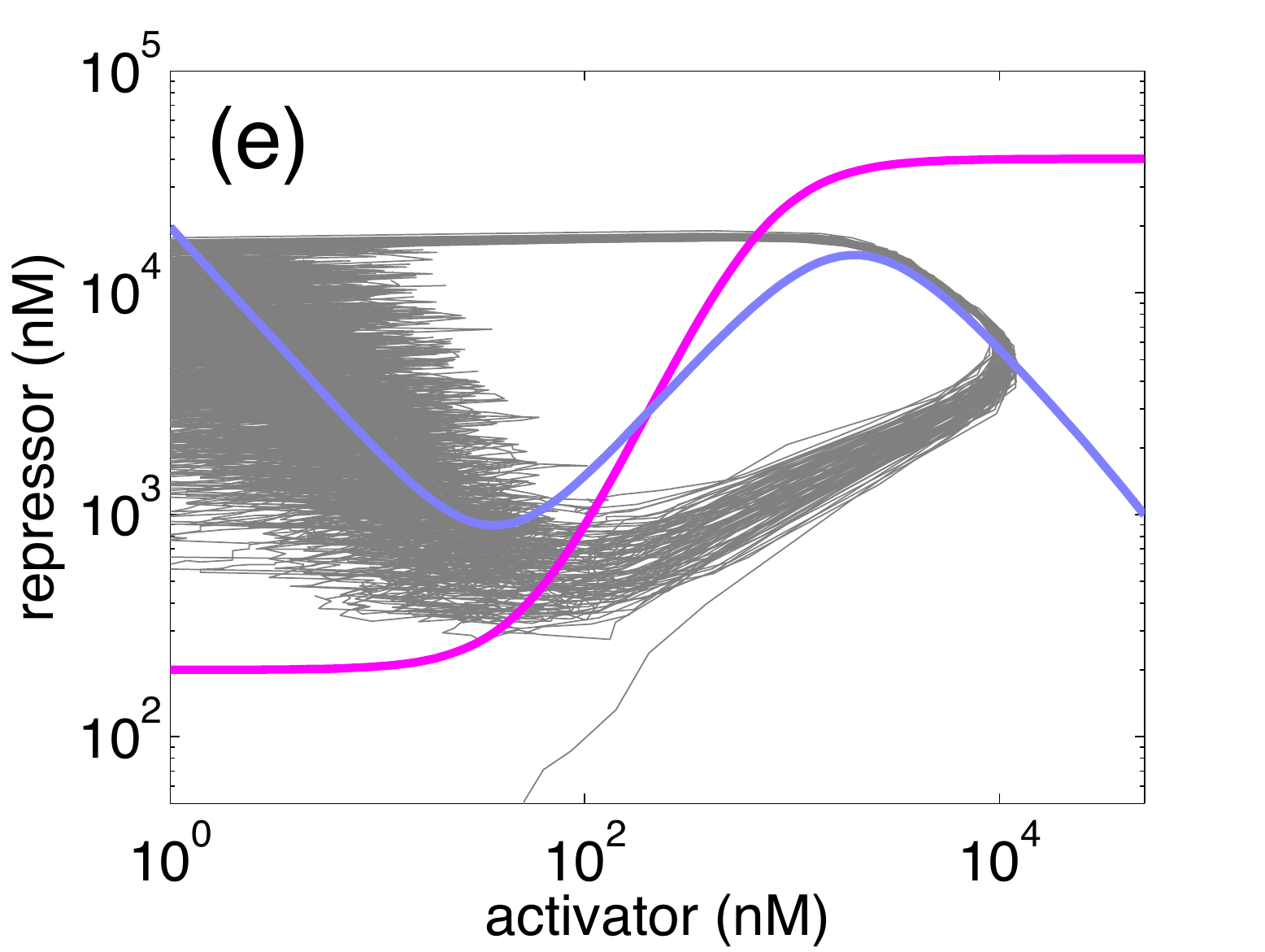}}}
\centerline{\resizebox*{14cm}{!}{\includegraphics{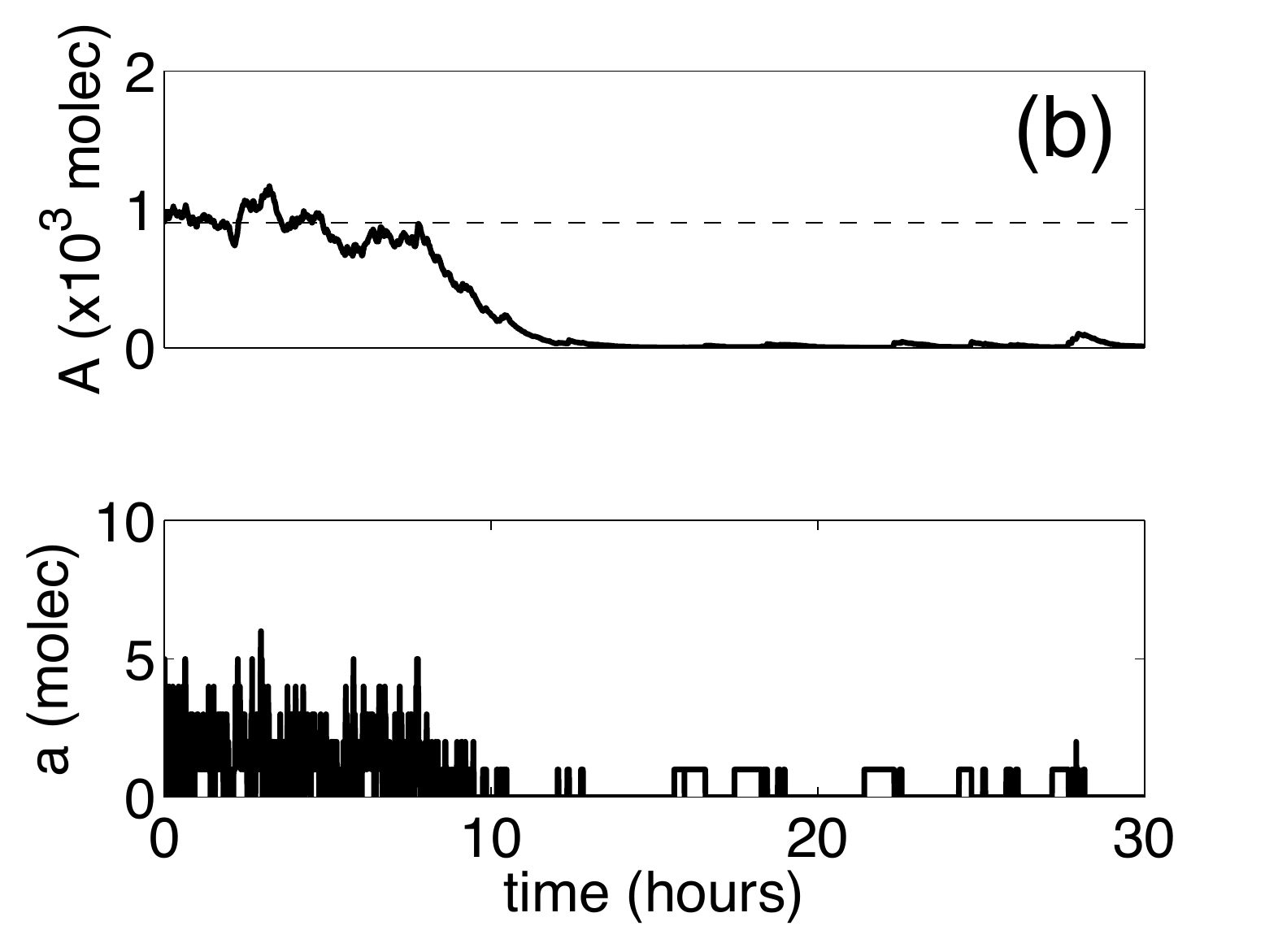}
\includegraphics{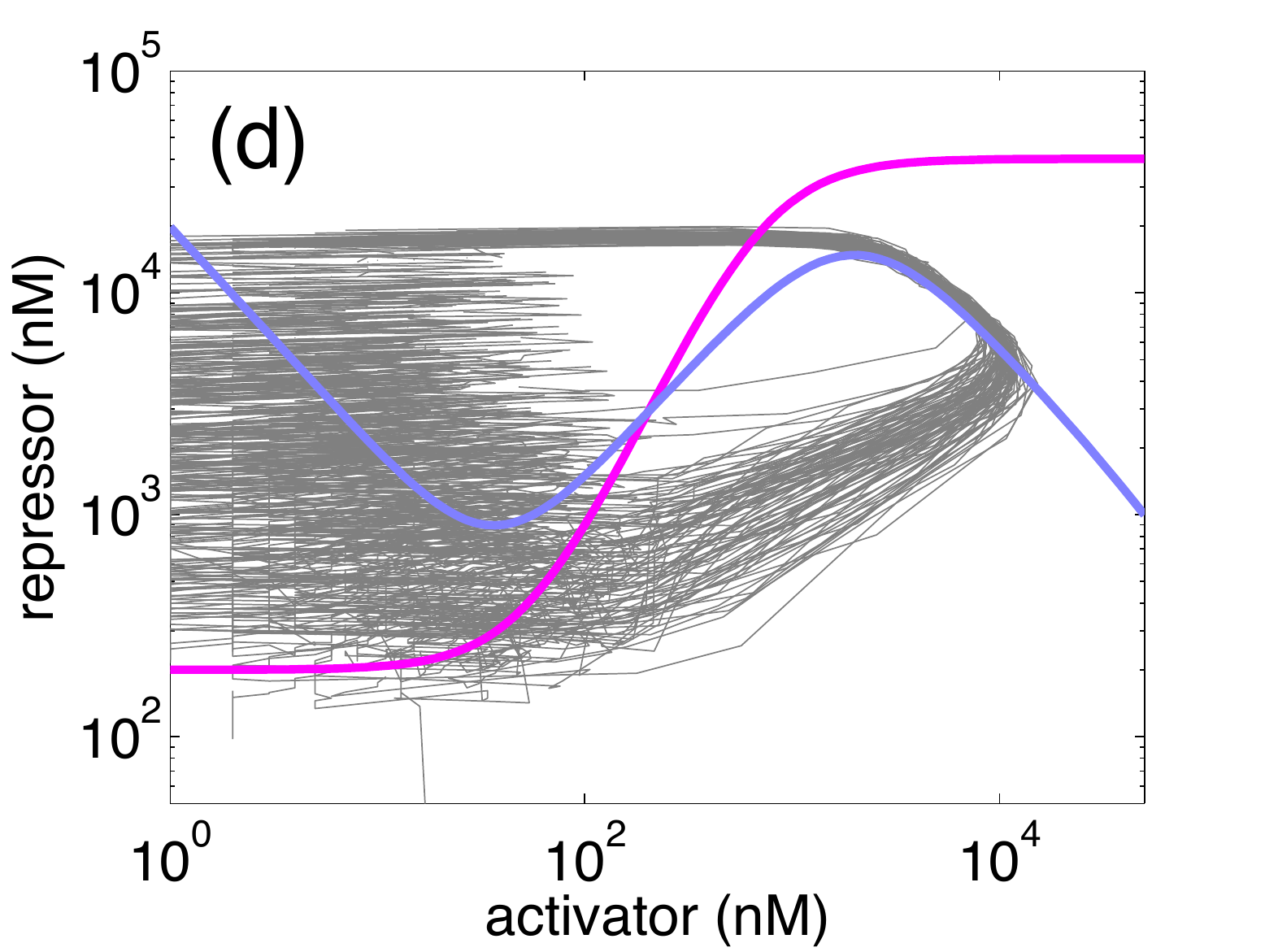}
\includegraphics{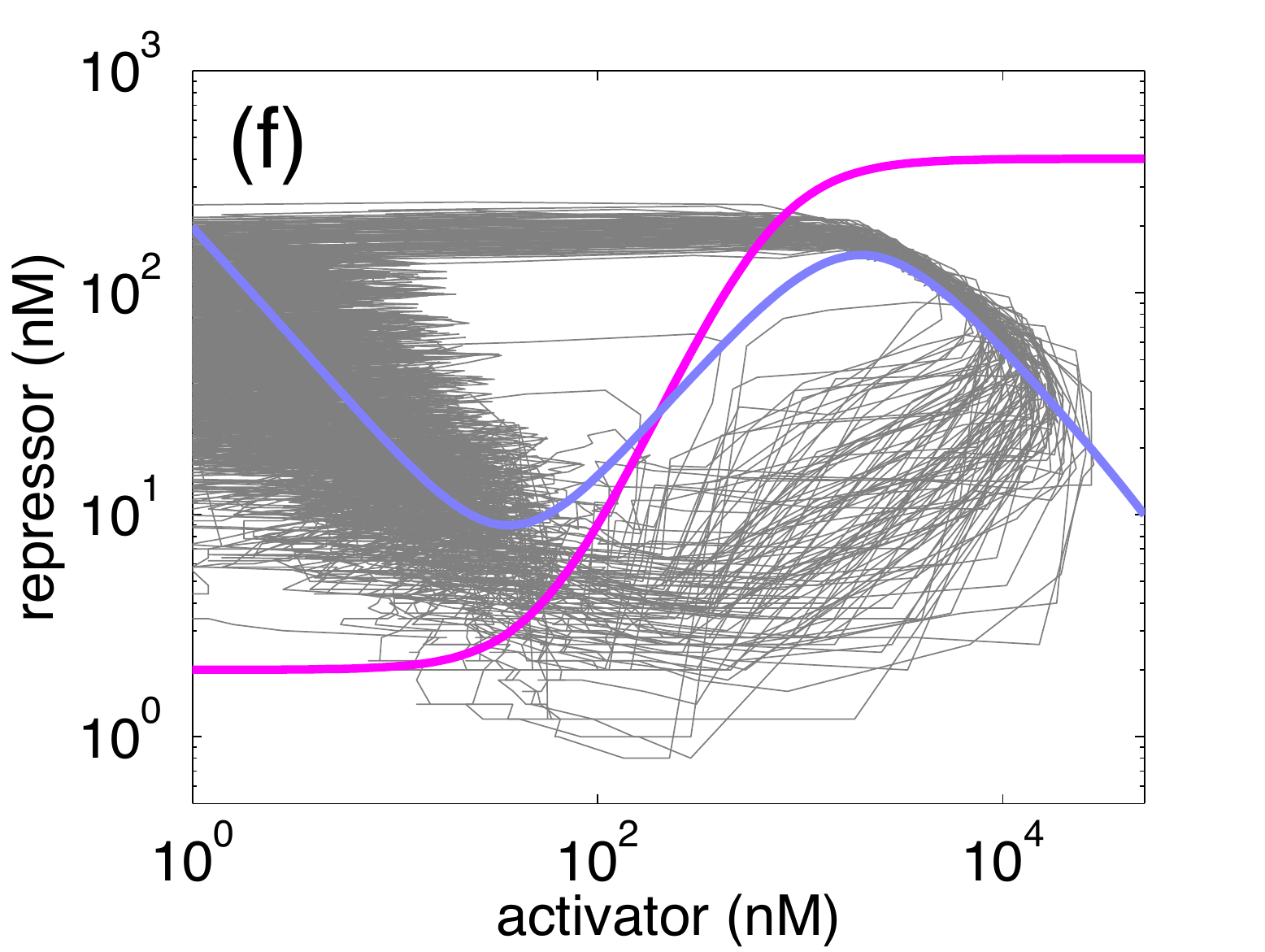}}}
\caption{Methods to control noise in genetic circuits. The left column
corresponds to a local method in which the transcription and translation
rates are simultaneously and inversely varied. The middle column represents
a global control of the noise that can be implemented by changing the cell
volume. The right column corresponds to a circuit-level version of the local
model, in which two interactions of the circuit are simultaneously and inversely
varied. The upper row shows a schematic representation of the three methods,
with the green arrows representing the direction in which the different regulation
processes (transcription, translation, and enzymatic degradation) have to vary for
noise to increase (while the average dynamics is unchanged).
Plots (a) and (b) show time series of the mRNA and protein levels for
increasing noise, and
plots (c-f) represent typical phase portraits (again with noise increasing going
downwards), with the activator and repressor
nullclines shown in pale blue and magenta, respectively.
Note that plots (c) and (e) are identical, representing the low-noise benchmark
for plots (d) and (f). Light dark lines
denote typical stochastic trajectories.
}
\label{fig:noise}
\end{figure}

In order to determine the effect of noise on the behavior of a given genetic circuit,
it would be desirable to control its intensity in experiments. One way of accomplishing this relies
on the interplay between the transcription and translation rates of a given gene.
If the transcription rate is low and the translation rate is high, mRNA molecules will
be created very infrequently, and each one of them will produce a burst of
protein expression. This is the translational bursting regime shown in
Fig.~\ref{fig:stoch}(b). In the opposite limit, if the transcription rate is high and
the translation rate is low, the number of mRNA molecules will be large, and
the associated fluctuations will be small, which will in turn lead to small
variations in the protein level. Thus, controlling independently (and inversely)
the rates of transcription and translation allows us to alter the amplitude of
the fluctuations without altering necessarily the mean expression level of the
protein of interest \cite{Ozbudak:2002p1237}. This situation is shown in Figs.~\ref{fig:noise}(a,b)
by means of numerical simulations of a simple positive feedback circuit. Figure~\ref{fig:noise}(a)
corresponds to the case of strong transcription and weak translation,
whereas the opposite case is shown in Fig.~\ref{fig:noise}(b). Here the
translational bursting is not so evident, since the circuit is operating in
the upper state of a bistable regime. In fact, for the situation of Fig.~\ref{fig:noise}(b)
the noise level is large enough to induce a jump of the circuit from the upper to the lower
state, which corresponds to the circuit turning off after approximately 10 hours.
In order to apply this method experimentally, it is necessary to mutate the promoter region
so as to alter the transcription rate, while at the same time mutating the
ribosome-binding site (RBS, precursor of the mRNA region where the ribosome
binds in order to start translation), which changes the translation rate
\cite{Ozbudak:2002p1237}. The technique has been recently applied
to show that initiation of competence in {\em B. subtilis} is stochastic \cite{Maamar:2007p619}.

The previous method is local, in the sense that only the noise acting upon a
given gene can be tuned. Furthermore it requires performing taylored mutations
to the promoter and RBS for each gene whose noise is to be varied. A more
straightforward way of controlling the noise would be to do it globally in the cell,
for all genes and molecular species. This relies on the fact that scaling up the
number of all components in the cell would leave all average quantities unchanged
and only change the noise. In that way the limit of infinite cell size would be a
sort of {\em thermodynamic limit}. An example of this is
shown in Figs.~\ref{fig:noise}(c,d). Experimentally, this can be accomplished
in bacteria by using mutations that prevent the cells from dividing, while they
are still growing and replicating their chromosome (which is necessary in order to scale up the whole
process). In that way, a ten-fold increase in the cell volume, accompanied by a
ten-fold increase in the copy number of the promoters from which all genes are
expressed (so that eventually or mRNAs and proteins are scaled up by the same
factor), and a ten-fold reduction in the bimolecular reaction rates (which
depend on the volume \cite{Gillespie:1977p456}) would lead to a ten-fold
decrease in the amount of noise acting upon the cell. This method has been
used in {\em B. subtilis} to prove that the initiation of competence is stochastic
\cite{Suel:2007cr}: as the cells grow in size (without dividing) the noise should
decrease and the probability of competence should decrease, as was
found experimentally. More recently, this technique has been used to
show that Min pole-to-pole oscillations in dividing {\em E. coli} cells
become more regular the longer the cell is, which indicates that
the regulation of Min activity is cell-length dependent \cite{FischerFriedrich:2010p952}.

Finally, in certain situations we might be interested in probing how noise in a
certain {\em interaction} of the circuit affects the behavior of the system. To that end,
we would need to perturb the circuit in such a way that the average effect of the
interaction being probed does not change, but only its fluctuations
[see Figs.~\ref{fig:noise}(e,f)]. To do that,
one can adapt the local method described above, by combining a promoter
mutation that alters the transcription rate of one species in the circuit, with
another mutation (in a binding site, for instance) that reduces the activity
of the resulting protein on its target by the same amount. In that case,
we alter the amount of noise acting upon that link without changing
the average effect on the other system component. This method has been
recently used, again in {\em B. subtilis}, to show that the high variability
in the durations of competence events is due to the low number of ComS
molecules existing during competence, and thus to the large noise in
the interaction on which ComS is acting \cite{Cagatay:2009kx}.

\subsection{Using noise}

Most of the studies devoted over the years to investigating the
effect of noise in the dynamics of genetic circuits have concentrated on
the robustness of the cell's behavior to the presence of noise
\cite{Hasty:2000p1290,Gonze:2002p1214,Chabot:2007p819,DiTalia:2007p411,Kar:2009p781,Vilar:2002p1289}.
Theoretical studies, on the other hand, have indicated the possibility that noise
induces new states, either stationary or oscillatory
\cite{Hasty:2000p1290,Vilar:2002p1289,Steuer:2004p613,Samoilov:2005p1250}, or enhance
information processing in the cell \cite{Paulsson:2000p1098}. We have recently proposed,
for instance, that intrinsic noise is able to stabilize an unstable steady state, leading
to a quantized distribution of excursion times \cite{Turcotte:2008ve} similar to those found experimentally
in the cell cycles of various species \cite{Klevecz:1976p829,Sveiczer:1996p662,Masui:1998p581}.
In a different context, signaling networks are subject to numerous sources of noise, both internal
and external to the cell. It is thus tempting to expect that cells would use that noise for their
own benefit, and in fact a recent study of a Boolean signaling network has shown that
environmental noise external to the cell in the form of background chatter, is able to improve
the response of the network to temporally structured input signals \cite{Rue:2010fk}.

Recent experimental evidence of the beneficial effects of random fluctuations
in genetic circuits regards for instance the response of cells to fluctuating and
uncertain environments \cite{Thattai:2004p541,Kussell:2005p1134,Acar:2008p923,Cagatay:2009kx}.
Another situation in which noise plays a useful role is in facilitating developmental evolution
through partial penetrance of mutants \cite{Eldar:2009p1294,Raj:2010p993}. In the next
years we should expect to see increasingly more examples of how cells use noise,
which is otherwise unavoidable, in a constructive manner.

\section{Multicellular dynamics}\label{sec:coupling}

Cells, both in unicellular and multicellular organisms, possess multiple mechanisms of 
communicating with one another, in a process that is critical for the survival of any
species. From bacterial biofilms to the more sophisticated animal tissues, cells
coordinate their behavior in order to function, and dynamics is not an exception.
Given the ubiquity of biochemical oscillations, discussed in Sec.~\ref{sec:osc} above,
one of the first questions that arose in this area is whether a collection of cells with
intrinsically dynamic behavior could synchronize with one another \cite{Winfree:2001uq}.
The question was biologically important, since organs such as the suprachiasmatic
nuclei in the mammalian brain, location of the circadian clock, were composed of thousands
of cells that had to operate synchronically in order to produce a single and clear
rhythm \cite{Liu:1997kx}. The nature of the synchronization mechanism in
multicellular circadian clocks is still under debate, although theoretical studies have
shown that such synchronization is feasible \cite{Gonze:2005vn}, even under the presence
of random fluctuations in the illumination level, which can in fact help the cells acquire
a rhythm even when individually they are arrhythmic \cite{Ullner:2009ys}.

Prior to these studies, numerical simulation results have already shown the feasibility
of synchronizing synthetic genetic oscillators by coupling the cells
by means of a quorum sensing mechanism \cite{McMillen:2002ff,Garcia-Ojalvo:2004kl}.
Synchronization reduces the frequency variability of the individual clocks, thus
converting a collection of sloppy oscillators into a relatively precise clock. Synchronized
genetic oscillations via quorum sensing have been recently reported in a synthetic
system \cite{Danino:2010ss}. In that case, however, the genetic circuit that generates
the oscillation is mediated by the quorum sensing molecules, thus the individual cells do
not oscillate.

Depending on the polarity with which the coupling signal is introduced into the
cellular oscillator, other types of collective phenomena besides synchronization
may arise \cite{ullner}. This includes, for instance, multistability and clustering \cite{Ullner:2007qf},
chaos \cite{Ullner:2008ly}, and cooperative differentiation \cite{Koseska:2010uq}.
When coupled with cell growth, we can use the dependence of the noise
on the population size \cite{Tanouchi:2008p598} to induce differentiation,
understood as a population arrest, upon arrival at a certain cell density
\cite{Koseska:2009zr}.

Finally, we turn our attention to natural genetic circuits in which coupling
plays a relevant role. The most natural field in which this happens is
development, since developmental processes are intrinsically dynamical and multicellular
\cite{Oates:2009p755}.
In fact, one of the more robust signalling oscillations is exhibited by
Hes1, a signaling molecule of Notch, which is one of the major mediators
of intercellular coupling. Hes1 oscillates with a period of around 2~hours, of the
same order as the oscillations of NF-$\kappa$B and STAT, both in culture \cite{Hirata:2002p353}
and {\em in vivo}. In this latter case, Hes1 oscillations are an important feature
of somite segmentation in the vertebrate embryo. This process is driven by
the segmentation clock, a collection of cells located in the presomitic
mesoderm, which are coupled to each other via Notch-Delta signaling.
In an elegant study, the strength of the coupling between the cells was
varied in a controlled way by chemical means, in order to see whether the
oscillations disappeared altogether, or if the cells simply became desynchronized.
The experiments unveiled a clear synchronization transition controlled by
the coupling strength \cite{Riedel-Kruse:2007tw}.

\section{Conclusion}

Gene regulation is a noisy dynamical nonlinear process. These characteristics are
frequently encountered by physicists in their studies of inanimate matter, thus it
seems a natural step to apply the techniques used in physics to the
investigation of living systems. In spite of the complexity of established genetic
networks, dynamical behavior in cells allows us to expect that smaller genetic
circuits can be associated with specific functionalities. These genetic circuits
can be studied with standard methods from the fields of nonlinear and statistical physics.
Here we have presented an overview of some of these techniques, both theoretical
and experimental, described some of the experimental evidence of dynamical
phenomena in genetic circuits, and discussed how such phenomena are used to
constrain and establish genetic circuits that explain cellular behavior.

\section*{Acknowledgments}

I thank Michael Elowitz, G\"urol S\"uel, Alfonso Mart\'inez-Arias, and the researchers
in their labs, for countless discussions that have taught me all that is written here.
I would also like to thank the members of my research group and my long-time
collaborators Alexey Zaikin, Evgeny Volkov, Aneta Koseska, and
J\"urgen Kurths. Special thanks are due to N\'uria Domedel-Puig, Antonio J. Pons, and
Ekkehard Ullner for carefully reading the manuscript and offering insightful comments.
This work has been financially supported by the Ministerio de
Ciencia e Innovaci\'on (Spain, project FIS2009-13360), the Spanish Multiple
Sclerosis Network (REEM-ISCIII), the Generalitat de Catalunya (project 2009SGR1168),
and the ICREA Academia programme.

\bibliographystyle{tCPH}
\bibliography{physgrc_arxiv}

\label{lastpage}

\end{document}